\documentclass[9pt,twoside]{pnas-new}

\templatetype{pnasmathematics} 
\usepackage{siunitx}
\usepackage{booktabs}
\usepackage{caption}
\usepackage{subcaption}
\usepackage{hyperref}
\usepackage{colortbl}
\usepackage{mathtools}
\usepackage{nowidow}
\usepackage{pdflscape}
\usepackage{afterpage}

\title{How to Identify Suitable Gate Dielectrics for Transistors based on Two-Dimensional~Semiconductors}

\author[a,*]{Theresia~Knobloch}
\author[b]{Quentin~Smets}
\author[c]{Anton~E.~O.~Persson}
\author[a]{Pedram~Khakbaz}
\author[a]{Christoph~Wilhelmer}
\author[b]{Dennis~Lin}
\author[c]{Zherui~Han}
\author[d]{Yunyan~Chung}
\author[e]{Kevin~P.~O'Brien}
\author[e]{Chelsey~Dorow}
\author[f]{Cormac~Ó~Coileáin}
\author[g]{Mario~Lanza}
\author[a]{Dominic~Waldhoer}
\author[a]{Alexander~Karl}
\author[h]{Kailang~Liu}
\author[h]{Tianyou~Zhai}
\author[i]{Hailin~Peng}
\author[i]{Congwei~Tan}
\author[j]{Xiao~Renshaw~Wang}
\author[f]{Georg~S.~Duesberg} 
\author[k]{John~Robertson} 
\author[e]{Uygar~Avci} 
\author[d]{Iuliana~Radu} 
\author[c]{Eric~Pop} 
\author[b]{Cesar~J.~Lockhart~de~la~Rosa}
\author[a,*]{Tibor~Grasser}
\affil[a]{Institute for Microelectronics, TU Wien, Gusshausstrasse 27-29, 1040 Vienna, Austria}
\affil[b]{Imec, Kapeldreef 75, Heverlee 3001, Belgium}
\affil[c]{Stanford University, Stanford 94305, USA}
\affil[d]{Corporate Research, Taiwan Semiconductor Manufacturing Company (TSMC), Hsinchu 30075, Taiwan}
\affil[e]{Intel Corporation, Components Research, Hillsboro, Oregon, USA}
\affil[f]{Institute of Physics, University of the Bundeswehr Munich, Neubiberg 85577, Germany}
\affil[g]{Department of Materials, Science and Engineering, National University of Singapore (NUS), Singapore}
\affil[h]{School of Materials Science and Engineering, Huazhong University of Science and Technology, Wuhan, China}
\affil[i]{Center for Nanochemistry, Peking University, Beijing, China}
\affil[j]{School of Electrical and Electronic Engineering, Nanyang Technological University, 50 Nanyang Ave, 639798, Singapore}
\affil[k]{Engineering Department, Cambridge University, Cambridge CB2 1PZ, United Kingdom}
\affil[*]{knobloch|grasser@iue.tuwien.ac.at}
\leadauthor{Knobloch} 

\definecolor{myellow}{HTML}{E8C378} 
\definecolor{mgreen}{HTML}{8AD38A} 
\definecolor{molive}{HTML}{E8E778} 
\definecolor{mblue}{HTML}{56AFAF} 
\definecolor{mora}{HTML}{E8AC78} 

\colorlet{tableora}{mora!20!white} 
\newcommand{\roworange}{\rowcolor{tableora}}
\colorlet{tableyellow}{molive!20!white} 
\newcommand{\rowyellow}{\rowcolor{tableyellow}}
\colorlet{tablegreen}{mgreen!20!white} 
\newcommand{\rowgreen}{\rowcolor{tablegreen}}

\colorlet{tabblue}{mblue!30!white} 
\newcommand{\headcolor}{\rowcolor{tabblue}}

\begin{document}
\begin{abstract}

\end{abstract}
 \maketitle
\thispagestyle{firststyle}
\ifthenelse{\boolean{shortarticle}}{\ifthenelse{\boolean{singlecolumn}}{\abscontentformatted}{\abscontent}}{}

\vspace*{-3cm}
\section*{Abstract} 

The recent progress in nanosheet transistors has established two-dimensional~(2D) semiconductors as viable candidates for future ultra-scaled electronic devices.
Next to reducing contact resistance, identifying good gate dielectrics is a fundamental challenge, as the dielectric/channel interface dramatically impacts virtually all performance parameters.
While several promising gate dielectrics have recently been reported, the evaluation of their quality and suitability is often fragmentary and focused on selected important performance metrics of the gate stack, such as the capacitive gate control,
leakage currents, reliability, and ease of fabrication and integration. 
However, identifying a suitable gate stack is a complex problem that has not yet been approached systematically. 
In this perspective, we aim to formulate general criteria for good gate dielectrics.

\section{Introduction} 

The continued down-scaling of transistor dimensions over seven decades has enabled breathtaking technological revolutions.
In the last decade, this process has gradually slowed as device dimensions reach fundamental physical limits with few-atoms-thick channels, the so-called nanosheets~(NS). 
At the current state-of-the-art, leading semiconductor manufacturers use stacked NS transistor designs, where multiple thin silicon channels are placed on top of each other to maximize both gate control and current density through the ultra-short channels~\cite{Cao2023}. 
As the gate is wrapped around the nanosheet, gate all around~(GAA) and NS transistor refer to the same geometry. 
For future device generations, complementary-field-effect-transistors~(CFETs) are a promising option, where p-type transistors are stacked on top of n-type transistors or vice-versa, thereby reducing the space required for a complementary~MOS~(CMOS) cell by about 40\%~\cite{Kukner2024}. 
For silicon nanosheet thicknesses below $3-\SI{5}{nm}$, charge carrier scattering at the interfaces increases drastically, thereby degrading the mobility~\cite{Uchida2002, Agrawal2024}. 
In addition, the moderate density of states in silicon channels limits the quantum capacitance and consequently the achievable gate control in silicon NS transistors~\cite{Bennett2023}. 
These fundamental scaling limits could be overcome by using two-dimensional~(2D) semiconductors like transition-metal dichalcogenides~(TMDs) as channel materials~\cite{Liu2021, Cao2023}. 
However, using 2D semiconductors as channels in ultra-scaled FETs brings tremendous challenges, among which the identification and deposition of a suitable high-$\kappa$ gate stack~\cite{Illarionov2020, Lin2023} is critical. 
While several studies have highlighted promising novel gate dielectrics~\cite{Illarionov2019, Li2020c}, others have focused on specific important gate stack performance metrics such as the reduction of gate leakage currents~\cite{Knobloch2021, Osanloo2021} or the improvement of 2D FET reliability~\cite{Illarionov2020, Knobloch2022a}. 
However, the identification of a suitable gate stack is a multifaceted problem and a top-level view is lacking.
This perspective seeks to formulate clear criteria in the search for good gate dielectrics,
aiming to distinguish intrinsic material limitations from fabrication-related issues that could be resolved. 

Conventionally, prototype 2D FETs employ combinations of amorphous oxides \textendash{} HfO$_2$, Al$_2$O$_3$ and SiO$_2$ \textendash{} to form gate stacks~\cite{Chung2022, Dorow2022}, 
largely due to material availability and suitability for process integration.
Section~\ref{sec:fabrication} gives an overview of the possible fabrication methods as well as their limitations, evaluated for two different applications for 2D FETs, the first one being planar devices in the back-end-of-line~(BEOL) for power delivery, see Figure~\ref{fig:fabrication_FET}(a), the second one being 2D complementary FETs~(CFETs) in the front-end-of-line~(FEOL), see Figure~\ref{fig:fabrication_FET}(b)-(e), requiring the deposition of the insulator on both the top and the bottom of the 2D NS. 
In Section~\ref{sec:performance}, the performance, variability and reliability criteria for gate stacks are formulated, highlighting how critical performance aspects of 2D NS FETs are defined by the gate stack, see Figure~\ref{fig:perfeval}(a). 
Gate stacks need to provide excellent capacitive gate control, while minimizing gate leakage currents and phonon scattering of charge carriers.
To limit variability, high-quality interfaces and well-defined threshold voltages are required.
To increase reliability border trap densities should be minimized and thermal conductivity maximized. 
The developed criteria will facilitate future high-throughput computational searches for suitable combinations of gate dielectrics with 2D channels~\cite{Klinkert2020, Kumar2024}, without an over-reliance on gate leakage values~\cite{Osanloo2021}. 
To reassess the potential of the various novel dielectrics that have been recently suggested for their use with 2D channels, we provide 
an overview of dielectric candidates within gate stacks of 2D FETs in Section~\ref{sec:insulators}, see Figure~\ref{fig:insulators}. 
Finally, the performance of these insulators with respect to the defined metrics are summarized, and we 
identify which challenges are the most critical for enabling energy-efficient highly-scaled 2D NS FETs in commercial applications.

\section{Fabrication Methods}
\label{sec:fabrication}
\begin{figure}[!ht]
\includegraphics[width=\linewidth]{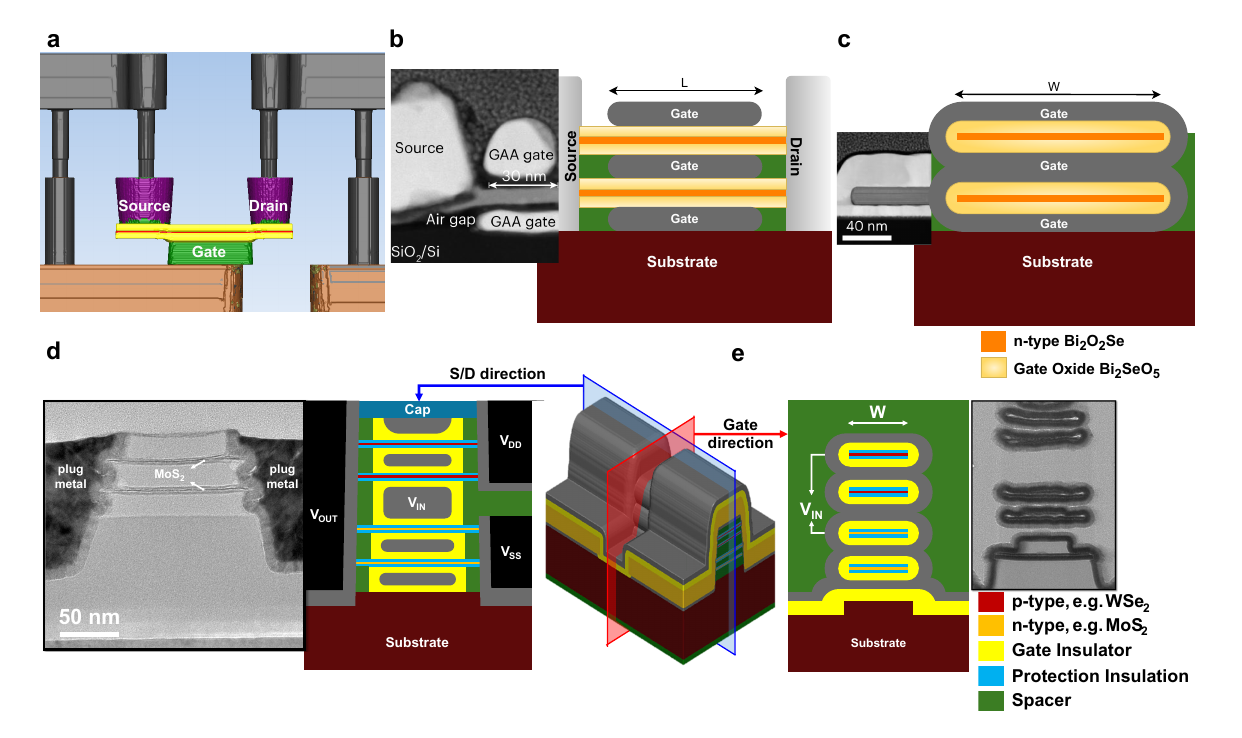}
\caption{
\textbf{(a)} Planar 2D FET with back gated and top contact configuration, integrated in the back-end or in the back-side power delivery network~(BSPDN), complementing high-performance silicon CMOS in the front-end.
\textbf{(b)-(c)} Schematic drawing of a stacked 2D GAA FET based on native oxides in \textbf{(b)} front view along the channel length and \textbf{(c)} side view along the channel width. The TEM images show the 2D GAA FET reported by~\cite{Tang2025}.
\textbf{(d)-(e)} Schematic drawing showing a stacked 2D GAA FET with all layers and components in \textbf{(d)} front view along the channel length and \textbf{(e)} side view along the channel width. The TEM images show the 2D stacked GAA FET reported by~\cite{Chung2023} in \textbf{(d)} and by~\cite{Chung2024} in \textbf{(e)}. Here, $V_\mathrm{DD}$ is the supply voltage, $V_\mathrm{SS}$ is common ground, $V_\mathrm{IN}$ and $V_\mathrm{OUT}$ are the input and output signals of a CMOS inverter realized within this CFET element.  }
\label{fig:fabrication_FET}
\end{figure}

The fabrication method for a gate stack depends on the 2D device configuration (back-, top-, double-gated, or GAA) and the intended application. One of the earliest entry points for 2D FETs in the roadmap~\cite{IRDS2023} is within the back-side power delivery network~(BSPDN), where they would act as power switches, bringing parts of the circuit into standby mode to reduce static power consumption~\cite{Lockhart2024}.
These switches are currently implemented in the FEOL and relocating them to the BSPDN would free up space on the costly wafer front side. Here, a simple planar back-gate or top-gate configuration would be sufficient, see Figure~\ref{fig:fabrication_FET}(a). 
The most cost competitive to silicon hybrid bonding would be monolithic deposition of the 2D channel and dielectric where the thermal budget must be BEOL-compatible (processing temperatures $\leq \SI{400}{\degree C}$). 
For more advanced technology nodes, 2D materials could outperform silicon in the CFET configuration~\cite{Ahmed2020a, Liu2021}, see Figure~\ref{fig:fabrication_FET}~(b-e), which poses additional challenges for insulator deposition, see Subsection~\ref{sec:fabrication}\ref{subsec:CFETs}. 
In general, there are two primary approaches: direct synthesis of a gate insulator on a 2D semiconductor, including via atomic layer deposition~(ALD)~(Subsection~\ref{sec:fabrication}\ref{subsec:ALD}), oxidation~(Subsection~\ref{sec:fabrication}\ref{subsec:nativeox}), or evaporation~(Subsection~\ref{sec:fabrication}\ref{subsec:epitaxy}); and transfer methods that stack the insulator and channel~(Subsection~\ref{sec:fabrication}\ref{subsec:transfer}).

\subsection{Insulator Deposition Requirements for 2D CFETs}
\label{subsec:CFETs}
Design-technology co-optimization~(DTCO) simulations indicate that 2D channels could outperform silicon in the ultra-scaled CFET configuration~\cite{Ahmed2020a}.
A prototype 2D CFET device layout, adapted from Chung \textit{et al.}~\cite{Chung2023, Chung2024}, is shown in Figure~\ref{fig:fabrication_FET}(d) along the source-drain direction and in (e) along the gate direction. 
To fabricate 2D CFETs in a gate last process, first a sacrificial insulator is deposited, then the protection insulation and the 2D channel are deposited monolithically, followed by the sacrificial top insulator.
These steps repeat until the entire 2D channel stack is grown, which is then etched into pillars, around which source and drain contacts are formed.
Next, all sacrificial layers are selectively etched, releasing the 2D channels, which remain protected by the protection insulation layers. 
Finally, the gaps between channels are filled with a high-$\kappa$ gate insulator using ALD or oxidation and gate metal deposition.

A key constraint for all direct growth methods is the thermal budget for insulator deposition.
Depending on the surrounding atmosphere, uncapped TMDs experience chalcogen out-gasing at temperatures above \SI{250}{\degree C} for WS$_2$~\cite{Wang2025} or above \SI{350}{\degree C} for MoS$_2$~\cite{Wang2025}. 
Once encapsulated, TMDs can typically withstand temperatures of up to \SI{550}{\degree C}~\cite{Zou2024, Wang2025}, even though the yield may suffer for anneals above \SI{350}{\degree C}.

\subsection{Atomic Layer Deposition}
\label{subsec:ALD}
ALD is the standard commercially applied method to deposit amorphous HfO$_2$ or dipole interlayers like Al$_2$O$_3$ or La$_2$O$_3$ in scaled Si technologies.
ALD offers conformal deposition~\cite{Zou2014, Wirtz2015}, allowing the coating of suspended channels in a CFET design. 
Yet, seeding an ALD layer on a defect-free van der Waals~(vdW) surface is challenging~\cite{Zou2014, Price2019}. 
Direct ALD of amorphous oxides on vdW surfaces nucleates poorly~\cite{Zou2014, Wirtz2015}. 
Plasma enhanced ALD~(PEALD) has been suggested to improve nucleation, but the plasma exposure damages the TMD layer, rendering it unsuitable for ALD growth on monolayer channels~\cite{Price2019}. 
A potentially better approach is to use organic seed layers~\cite{Park2016f, Li2019}, notably perylene-tetracarboxylic dianhydride~(PTCDA)~\cite{Li2019}, although most organic seed layers are thermally unstable above \SI{250}{\degree C}, rendering them CMOS incompatible. 
Alternatively, thin inorganic layers ($<\SI{1}{nm}$) can be evaporated on top of the TMD channel and subsequently oxidized before ALD of the gate dielectric, resulting in thin SiO$_2$, Al$_2$O$_3$, or Y$_2$O$_3$ interlayers.
However, evaporation is directional and unsuitable for GAA FETs~\cite{Jiang2023a, Ko2025a}. 
Under industry-compatible conditions, ALD gate stacks have been grown on TMDs by depositing a sub-nm dielectric seeding layer in a surface physisorption-based soaking approach using trimethylaluminum~(TMA) as precursor, resulting in a thin AlO$_x$ interlayer~\cite{Lin2020, Smets2021a}. 
Triisobutylaluminum~(TIBA) may offer better adhesion for physisorption on MoS$_2$ than TMA~\cite{Cho2025}, and triethylaluminum~(TEA) has recently been used to deposit an AlO$_x$ interlayer, followed by promisingly thin and uniform HfO$_2$~\cite{Ko2025}. 
Still, whether physisorption-based methods can deposit interlayers with low trap densities, e.g. below $\SI{e12}{cm^{-2}eV^{-1}}$, remains unclear, see Subsection~\ref{sec:performance}\ref{subsec:interface}.
Notably, PEALD was used for vdW epitaxy of crystalline hexagonal AlN (hAlN) on TMDs~\cite{Wang2023}, even though plasma damage of the TMD is a concern.
Despite considerable progress, none of the interlayers for HfO$_2$ ALD can currently offer a clean vdW interface with the 2D channel, as all methods introduce a relatively high defect density, either forming a defective interlayer, e.g., AlO$_x$~\cite{Smets2021a, Ko2025} or SiO$_2$~\cite{Ko2025a}, or creating defects in the TMD~\cite{Price2019, Wang2023}, even though the trap densities are gradually being reduced~\cite{Ko2025a}, see also Section~\ref{sec:performance}\ref{subsec:interface}. 
As interface defect densities decrease with improved growth methods, ALD nucleation will likely become more challenging.

\subsection{Oxidation}
\label{subsec:nativeox}

Oxidizing layered semiconductors into their respective native oxides yields high-quality interfaces and conformal coatings for complex geometries, including suspended channels in CFET designs.
A main challenge is to selectively oxidize only the top and bottom layers, without degrading the central semiconducting layer. 
X-ray photoelectron spectroscopy~(XPS) can monitor the oxidation thickness, but may lack sufficient sensitivity for industrial applications.
Depending on a metal's oxygen affinity, the oxide may deoxidize at elevated temperatures, requiring reliability evaluations at BEOL thermal budgets, see Subsection~\ref{sec:performance}\ref{subsec:hysteresis}. 
Layered semiconductors with native high-$\kappa$ oxides include ZrS$_2$ and ZrSe$_2$ (forming ZrO$_2$~\cite{Jin2023}), and HfS$_2$ and HfSe$_2$ (forming HfO$_2$~\cite{Mleczko2017, Kang2024}). 
To prevent over-oxidation, a trilayer HfSe$_2$/MoS$_2$/HfSe$_2$ stack could be used, where the outer layers of the vdW stack would be converted into HfO$_2$.
In addition, 2D FET prototypes have used the n-type semiconductor bismuth oxyselenide, Bi$_2$O$_2$Se~\cite{Wu2017, Tan2025}, which oxidizes into bismuth oxoselenate, Bi$_2$SeO$_5$. Annealing Bi$_2$O$_2$Se at about \SI{400}{\degree C} in an oxygen-rich environment~\cite{Li2020c, Tu2020} results in the alpha phase, while UV-assisted oxidation at about \SI{100}{\degree C} produces the crystalline beta phase via a layer-by-layer intercalative mechanism~\cite{Zhang2022, Tan2023}. 
Because Bi$_2$O$_2$Se can grow horizontally or vertically, its oxidation allows planar FETs~\cite{Zhang2022}, FinFETs~\cite{Tan2023} or GAA FETs~\cite{Tang2025}, see Figure~\ref{fig:fabrication_FET}~(b-c). 
To date, demonstrated Bi$_2$O$_2$Se transistors were based on multi-layer channels, and above \SI{150}{\degree C} gate leakage may increase due to possible deoxidation~\cite{Khakbaz2025}, see Subsection~\ref{sec:performance}\ref{subsec:leakage}. 
 

\subsection{Directional Deposition Methods}
\label{subsec:epitaxy}
Directional deposition methods, like molecular beam epitaxy~(MBE), thermal evaporation, and sputtering, are limited to back-gated, top-gated, and double-gated device designs, potentially targeting BEOL applications.
Epitaxial growth of crystalline insulators is theoretically possible on clean surfaces of TMDs, despite a potentially large lattice mismatch, in a vdW epitaxy~\cite{Koma1992}.
However, in practice, only a few lattice matched systems have been realized, growing for example thin crystalline calcium fluoride~(CaF$_2$) on silicon~\cite{Illarionov2019} or silicene~\cite{Nazzari2022}.
Major drawbacks of MBE growth are the required high vacuum and the slow growth rates.
These issues can be avoided with thermal evaporation, which, however, usually results in amorphous layers~\cite{Meng2024}.
Recently, superionic, amorphous, rare-earth metal fluoride films, including for example lanthanum fluoride~(LaF$_3$), have been evaporated on top of TMDs at room temperature~\cite{Meng2024}.

\subsection{Transfer of Dielectrics} 
\label{subsec:transfer}
Many dielectrics require either high synthesis temperatures above \SI{400}{\degree C} or growth substrates of a defined crystallinity, and thus cannot be grown directly on the 2D channel. 
Such dielectrics are typically grown separately and require transfer processes to be integrated into gate stacks.
While transfer is a powerful tool to investigate novel dielectrics in exploratory studies, the transfer of dielectrics is unlikely to be adopted by industry due to the increased cost and reduced yield of a considerably longer process flow. 
As such, most transfer processes are not scalable. 
Because most transfer methods rely on polymer carrier scaffold layers~\cite{Frisenda2018}, transferred dielectrics are often contaminated by particles from the scaffold and growth substrate~\cite{Tilmann2023}.
To avoid residues, polymer-free transfer methods were developed, such as silicon nitride cantilevers~\cite{Wang2023a} or the vdW pick-up transfer method~\cite{Frisenda2018}.
Recently, scalable transfer methods~\cite{Ghosh2024} for CVD-grown TMDs have been developed, even though they have not yet been used to transfer dielectrics. Independently, a wafer-scale transfer of ALD-grown amorphous Al$_2$O$_3$ and HfO$_2$ layers has been demonstrated, albeit with considerable variability~\cite{Lu2023}. For more details on insulator transfer see Section~SI\ref{subsec:SI:transfer}. 

\section{Performance, Variability, and Reliability Requirements}
\label{sec:performance}

The gate stack needs to satisfy numerous criteria, see Figure~\ref{fig:perfeval}(a), each relying on several material properties, see Figure~\ref{fig:perfeval}(b). 
Primarily, the insulator needs to ensure good gate control, see Subsection~\ref{sec:performance}\ref{subsec:capacitance}, while maintaining low leakage, see Subsection~\ref{sec:performance}\ref{subsec:leakage}, and minimizing insulator phonon coupling to the 2D channel, preserving high carrier mobilities, see Subsection~\ref{sec:performance}\ref{subsec:mobility}. 
These performance criteria define on- and off-current specifications.
Planar 2D-based power switches in the BSPDN~\cite{Lockhart2024} would require on-state resistances of about \SI{1}{\kilo \ohm}\textmu m at $V_\mathrm{G}=V_\mathrm{DD}$, $V_\mathrm{D}=0.01V_\mathrm{DD}$ (linear regime) and $I_\mathrm{on}/I_\mathrm{off}>10^5$.
This resistance is higher than for their silicon front-end counterparts; therefore 2D devices would need 10$\times$ more die area to deliver the required load current. 
This would be no problem given the available area on the wafer backside, but a careful trade-off is needed between the dynamic power dissipation caused by the increased gate capacitance and the energy savings enabled by the power switching. 
For CFETs, the criteria are quantified by the international roadmap for devices and systems~(IRDS)~\cite{IRDS2023}, see Table~\ref{tab:IRDS}. 
Importantly, FET performance should be evaluated at the IRDS-specified operating conditions.
For example, for the $\SI{5}{\AA}$ high density node, the on-current $I_\mathrm{D,on}$ in saturation must be evaluated at $V_\mathrm{G}=V_\mathrm{D}=V_\mathrm{DD}=\SI{0.6}{V}$ (target \SI{587}{\micro\ampere\per\micro\meter}) and the off-current at $V_\mathrm{G}=\SI{0}{V},\:V_\mathrm{D}=V_\mathrm{DD}=\SI{0.6}{V}$ (target \SI{0.1}{\nano \ampere\per\micro\meter}).
For minimizing variability, the interface trap density should be small, see Subsection~\ref{sec:performance}\ref{subsec:interface} and the threshold voltage~($V_\mathrm{th}$) should be well-defined, see Subsection~\ref{sec:performance}\ref{subsec:threshold}. 
For achieving reliable operation, the border trap density in the insulator should be reduced, see Subsection~\ref{sec:performance}\ref{subsec:hysteresis}, and dielectric breakdown should only occur after long stress, see Subsection~\ref{sec:performance}\ref{subsec:TDDB}. 
Finally, high thermal conductivity is needed to avoid overheating, see Subsection~\ref{sec:performance}\ref{subsec:heat}.

\subsection{Capacitive Gate Control}
\label{subsec:capacitance}
To ensure good electrostatic control, minimal short-channel effects, and high on-currents in 2D NS FETs with gate lengths below \SI{12}{nm}, a high gate capacitance density~($C_\mathrm{G}$) is key, since $I_\mathrm{D,on}\propto C_\mathrm{G}$. 
$C_\mathrm{G}$ is given by the series combination of the gate insulator capacitance density~($C_\mathrm{ins}$), the quantum capacitance density of the 2D semiconductor~($C_\mathrm{sc}$)~\cite{Bennett2023}, and the vdW gap capacitance density, $C_\mathrm{vdW}$, $C_\mathrm{G} = \left(C_\mathrm{sc}^{-1} + C_\mathrm{vdW}^{-1} + C_\mathrm{ins}^{-1}\right)^{-1}$.
Here, $C_\mathrm{ins} = \varepsilon_0 \varepsilon_\mathrm{ins} /t_\mathrm{ins}$ holds, with the vacuum permittivity $\varepsilon_0$, the dielectric constant $\varepsilon_\mathrm{ins}$, and the insulator thickness $t_\mathrm{ins}$, see Figure~\ref{fig:perfeval}(b).
A common benchmarking metric for gate stacks is the capacitance equivalent thickness, $\mathrm{CET}\vcentcolon =  \left(\varepsilon_0 \varepsilon_\mathrm{SiO_2} \right)/ C_\mathrm{G} $. 
A related metric is the equivalent oxide thickness, $\mathrm{EOT} \vcentcolon = \left(\varepsilon_0 \varepsilon_\mathrm{SiO_2} \right)/ C_\mathrm{ins} = t_\mathrm{ins} \left(\varepsilon_\mathrm{SiO_2}/\varepsilon_\mathrm{ins}\right)$.
EOT and $\varepsilon_\mathrm{ins}$ can be experimentally obtained from capacitance-voltage measurements~($C_\mathrm{G}(V_\mathrm{G})$) on metal-insulator-metal~(MIM) structures, ideally with an insulator thickness series, see Figure~\ref{fig:perfeval}(c).
Typically, $\varepsilon_\mathrm{ins}$ depends on $t_\mathrm{ins}$ due to variations in the oxide density, the degree of crystallinity, and dead-layer effects at the interfaces~\cite{Stengel2006}. 
\textit{Meanwhile, CET must be extracted from $C_\mathrm{G}(V_\mathrm{G})$ measurements on dedicated metal-insulator-semiconductor~(MIS) structures, which include the vdW gap as well as impurities at the 2D-semiconductor/insulator interface.} 
According to the IRDS~\cite{IRDS2023}, the target for sufficient gate control is a CET of \SI{0.9}{nm} in technology nodes beyond 2030. 
Whereas for bulk semiconductors the MIS test structures are typically simple vertical two-terminal devices, 2D~semiconductors and other ultra-thin body fully depleted semiconductors require lateral edge MIS capacitors with multi-finger layout, see Figure~\ref{fig:perfeval}(d).
In the absence of a neutral bulk region in fully depleted thin film capacitors, the source of carriers is the metal-semiconductor junction at the edge of the device.
An overview of measurement methods to evaluate CET is provided in Section~SI\ref{subsec:SI:capmeas}. 
In junctionless 2D FETs, a rough approximation is $\mathrm{CET} \approx  \mathrm{EOT} + \SI{0.4}{nm}$, a correction of which $\sim \SI{0.1}{nm}$ is channel capacitance and $\sim \SI{0.3}{nm}$ accounting for the channel-gate stack vdW gap~\cite{Zhang2022}.
Consequently, in scaled technology nodes, EOT should amount to \SI{0.5}{nm} or less.
However, definitions based on EOT are inherently imprecise because  vdW gaps can range from 0.2\,nm to 0.4\,nm~\cite{Luo2022}.
In addition, 2D materials may show poor adhesion to the gate stack and possible delamination can result in increased CET values due to a small air gap.
Reaching a CET below \SI{0.9}{nm} seems unlikely with a gate stack thicker than \SI{3}{nm}, as this would require a uniform $\varepsilon_\mathrm{ins}$ higher than $\sim 23$ in such thin layers which is extremely challenging and has to the best of our knowledge not been reached.

\subsection{Gate Leakage Currents}
\label{subsec:leakage}
To ensure low static power consumption, the gate leakage current $I_\mathrm{G}$ must be small, since $P_\mathrm{stat} = I_\mathrm{G} V_\mathrm{DD}$.
$I_\mathrm{G}$ depends on the insulator thickness, the band gap and band offsets, the tunneling masses, the vdW gap, interface quality, and insulator defect density, see Figure~\ref{fig:perfeval}(b). 
Charge traps in the insulator increase leakage via trap-assisted tunneling~(TAT)~\cite{Schleich2022}, while the defect-free lower limit of the leakage current can be estimated using the Tsu-Esaki model~\cite{Tsu1973, Knobloch2021}, see Section~\ref{sec:perfpotential} and Section~SI\ref{subsec:SI:method_leak}. 
Comparing different dielectrics requires evaluating $I_\mathrm{G}$ at the same CET, which ties $I_\mathrm{G}$ to $\varepsilon_\mathrm{ins}$, see Section~\ref{sec:performance}\ref{subsec:capacitance}.
Measured gate leakage currents, at use conditions ($V_\mathrm{G} = V_\mathrm{DD}$, $V_\mathrm{D} = \SI{0}{V}$), must be below the low-power limit 
of $\SI{700}{mAcm^{-2}}=\SI{7}{nA}$\textmu$\mathrm{m^{-2}}$ with the CET being below \SI{0.9}{nm}~\cite{IRDS2023}, see Table~\ref{tab:IRDS}. 
To ensure technological readiness, $I_\mathrm{G}\left(V_\mathrm{G}\right)$ should be measured on several MIS structures with device areas spanning from 0.001\textmu$\mathrm{m^{2}}$ to 1\textmu$\mathrm{m^{2}}$ ~\cite{Wu2021a, Zeng2024}, see Figure~\ref{fig:perfeval}(c). 
In particular for more defective dielectrics, the measured leakage is often smaller on smaller areas, as larger areas may contain pinholes.
In addition, measurements should be performed at different temperatures, since this can be used to distinguish direct tunneling from TAT~\cite{Schleich2022}. 
Due to TAT, thickness variations, and surface roughness, leakage currents are a statistical quantity with considerable variance, requiring comprehensive statistical analysis.

\subsection{Impact on Semiconductor Mobility} 
\label{subsec:mobility}
Although the semiconductor mobility is often considered a material property of 2D semiconductors, especially in monolayers it strongly depends on the dielectric surrounding and thus on the gate stack~\cite{Ma2014, Gopalan2022}.
For example, encapsulating monolayer WSe$_2$ in thick hBN yields a room temperature hole mobility of up to \SI{840}{cm^2V^{-1}s^{-1}} ~\cite{Liu2023}, whereas typical devices reach less than \SI{100}{cm^2V^{-1}s^{-1}}~\cite{Movva2015, OBrien2021}.
The mobility is influenced by various scattering mechanisms, including impurity scattering and remote surface-optical~(SO)/ remote phonons, see Figure~\ref{fig:perfeval}(b).
The remote phonons originate from the polar phonon modes of insulators, which are particularly pronounced in insulators with strongly polarized structures, like HfO$_2$, indicating a large dielectric constant.
Also interface charges degrade the mobility via impurity scattering~\cite{Gopalan2022}, see Section~\ref{sec:perfpotential}, and Section~SI\ref{subsec:SI:method_mobility}.
Moreover, high strain or large electric fields have an effect on the separation between Q and K valleys, thereby increasing intervalley scattering and reducing the mobility. 
In most prototype 2D FETs, mobilities are limited by scattering at charged impurities, but if impurity densities are lowered in the future, remote SO phonon scattering will determine the ceiling of the attainable mobilities~\cite{Ma2014}. 
As a rule of thumb, interfaces with high-$\kappa$ dielectrics like HfO$_2$ reduce the semiconductor mobility the most, and usually higher dielectric constants correlate with lower semiconductor channel mobilities.
Moreover, phonon coupling will be more efficient for out-of-plane dipoles of quasi vdW interfaces (e.g. CaF$_2$), see Subsection~\ref{sec:insulators}\ref{subsec:fluorides}, or at defective interfaces, see Subsection~\ref{sec:insulators}\ref{subsec:amorphous}. 
Accurately measuring the channel mobility requires separating intrinsic mobility from contact effects, which can be achieved with dedicated test structures for transfer length measurements, four-probe measurements, or Hall measurements~\cite{Pang2021a, Cheng2022}, see Figure~\ref{fig:perfeval}(c).
The electron mobility relevant for the IRDS-defined on-current should be evaluated at the corresponding carrier concentration of $\sim \SI{e13}{cm^{-2}}$.

\subsection{Interface Trap Density}
\label{subsec:interface}
Both for the power switch and CFET applications, small interface trap densities are key,
because they impact the subthreshold swing, $SS = \mathrm{log}\left(10\right) \left(k_\mathrm{B}T/q\right) \left(C_\mathrm{ins}+C_\mathrm{sc}+q^2 D_\mathrm{it}\right)/C_\mathrm{ins}$, where $k_\mathrm{B}$ is Boltzmann's constant, $T$ temperature, $q$ the elementary charge and $D_\mathrm{it}$ the density of interface traps. 
For technology nodes beyond 2030, $SS$ should be smaller than \SI{65}{mV/dec.}~\cite{IRDS2023}, approaching the ideal value of \SI{59.6}{mV/dec.}. 
To achieve an on/off current ratio of \SI{e7}{} within a $V_\mathrm{G}$ window of $[0,V_\mathrm{DD}]$ (Table~\ref{tab:IRDS}), a small $SS$ is required in the entire subthreshold regime.
Even though most research groups are currently reporting $SS_\mathrm{min}$ at arbitrary current levels, the relevant quantity is $SS_\mathrm{avg}$, averaged over at least five orders of magnitude, see Figure~\ref{fig:perfeval}(c).
Often $SS$ is degraded at higher current levels due to Schottky contacts and higher $D_\mathrm{it}$ near the band edges.
So far, $SS_\mathrm{avg} \leq \SI{65}{mV/dec.}$ has been reached for long channel n-type FETs~\cite{Tang2025} and a few demonstrations of n-type FETs with gate lengths down to \SI{20}{nm}~\cite{Jiang2023a, Jiang2024}, see Table~\ref{tab:insulators}, while for p-type FETs, $SS$ is often worse due to a higher $D_\mathrm{it}$~\cite{Park2016f, OBrien2021}.
Short-channel effects degrade $SS_\mathrm{avg}$ for gate lengths below \SI{100}{nm}, hence $SS_\mathrm{avg}$ should be reported as a function of the gate length on short-channel devices.
In order to reach $SS_\mathrm{avg} \leq \SI{65}{mV/dec.}$ and to limit variability, $D_\mathrm{it}$ needs to be lower than $\SI{e12}{cm^{-2}eV^{-1}}$.
At the moment, $D_\mathrm{it}$ reported for prototype 2D FETs are in the range of \SI{e12}{}-$\SI{e14}{cm^{-2}eV^{-1}}$~\cite{Fang2018, Gaur2019}. 
These measured values include channel and interface traps and depend heavily on how they are characterized, since different methods capture traps with different time constants and energy levels. 
While $SS_\mathrm{avg}$ depends on $D_\mathrm{it}$, measurements of $SS_\mathrm{avg}$ cannot be used to properly determine $D_\mathrm{it}$, as $SS_\mathrm{avg}$ is a convoluted quantity. 
For analyzing $D_\mathrm{it}$, $C_\mathrm{G}\left(V_\mathrm{G}\right)$ measurements need to be performed on multi-finger MIS capacitors with areas on the order of 100-1000\textmu $\mathrm{m^{2}}$~\cite{Gaur2019, Gaur2020}, see Figure~\ref{fig:perfeval}(d) and Section~SI\ref{subsec:SI:measinttraps}.
Measurements at different frequencies $\SI{1}{kHz}-\SI{1}{MHz}$ and temperatures $\SI{4}{K}-\SI{300}{K}$ enable the extraction of $D_\mathrm{it}\left(E\right)$ profiles~\cite{Gaur2019, Mootheri2021}, see for example gate stacks on MoS$_2$ and WS$_2$ in Figure~\ref{fig:perfeval}(e). 
For both channels, $D_\mathrm{it}\left(E\right)$ near mid-gap is close to the targeted $\SI{e12}{cm^{-2}eV^{-1}}$, but near the band edges, $D_\mathrm{it}$ increases exponentially to $\SI{e14}{cm^{-2}eV^{-1}}$, causing a stretch-out of the transfer characteristics.
This stretch-out prevents low-power operation and remains a key gate stack challenge. 

\subsection{Threshold Voltage Variability}
\label{subsec:threshold}
$V_\mathrm{th}$ is determined by the charge balance in the channel and insulator, along with the gate metal's work function.
It depends on the dielectric constant, the interface quality, and the density of charged defects in the insulator, see Figure~\ref{fig:perfeval}(b).
In general, 2D FETs must operate in enhancement mode, ensuring the devices are off at $V_\mathrm{G} = \SI{0}{V}$. Thus, for n-type FETs $V_\mathrm{th}$ should be positive and for p-type FETs negative.
Yet, most 2D devices operate in depletion mode with normally-on 2D channels, see Table~\ref{tab:insulators}. 
Generally, minimizing defects, for example sulfur vacancies in MoS$_2$~\cite{Song2017a, Sarkar2019}, helps to avoid a negative $V_\mathrm{th}$ in nFETs~\cite{Smithe2017a,Lan2024}.
Additional $V_\mathrm{th}$ control can be achieved by introducing charges or dipoles at the interface~\cite{Ko2025, Ko2025a} or by selecting gate metals with suitable work functions. 
As 2D FETs mature, a precise tuning of $V_\mathrm{th}$ will be essential, allowing for multiple threshold voltages in a single technology, including for example $V_\mathrm{th}=\SI{0.26}{V}$~\cite{IRDS2023}, see Table~\ref{tab:IRDS}.
Moreover, device-to-device variation of $V_\mathrm{th}$ is a key metric. For sufficiently large sample sizes~($>$100 devices), the standard deviation of $V_\mathrm{th}$ should be reported as a function of device area~\cite{Shi2021}, to assess the impact of the gate stack on the variability, see Figure~\ref{fig:perfeval}(c). 

\subsection{Hysteresis and Drifts}
\label{subsec:hysteresis}
Hysteresis in the transfer characteristics and long-term $V_\mathrm{th}$ drifts are frequently observed issues in prototype 2D FETs~\cite{Late2012, Knobloch2022a}. 
These phenomena depend on the insulator thickness, dielectric constant, interface quality, morphology and potential ferroelectricity, see Figure~\ref{fig:perfeval}(b). 
Their primary microscopic origin is charge trapping at border traps in the gate insulator, which exhibit time constants ranging from nanoseconds up to years~\cite{Grasser2012}.
During a double $I_\mathrm{D}\left(V_\mathrm{G}\right)$ sweep, a subset of traps will capture charge during the up sweep but not emit during the down sweep, causing a hysteresis.
The hysteresis width is evaluated as $\Delta V_\mathrm{H}= \pm \left(V_\mathrm{th}^\mathrm{down} - V_\mathrm{th}^\mathrm{up}\right)$, with $V_\mathrm{th}^\mathrm{down}$ and $V_\mathrm{th}^\mathrm{up}$ denoting the threshold voltages of the down and up sweep, and $-$ for p-type FETs and $+$ for n-type FETs. 
The observed $\Delta V_\mathrm{H}$ will depend on the sweep time~($t_\mathrm{SW}$), temperature, and the sweep voltage range~\cite{Karl2025, Late2012}.
Hence hysteresis measurements should span orders of magnitude in sweep times and a wide temperature range, see Figure~\ref{fig:perfeval}(c) and Section~SI\ref{subsec:SI:meashystBTI}. 
Bias temperature instability~(BTI) measurements analyze the reliability of FETs on longer time scales.
In a BTI measurement, $V_\mathrm{G}$ is switched between two discrete levels (stress condition $V_\mathrm{H}$ and recovery state $V_\mathrm{L}$) and the drifts of $V_\mathrm{th}$ are measured.
$\Delta V_\mathrm{th}$ transient drifts are monitored at geometrically increasing intervals during both the stress and recovery phases, for example using very fast $V_\mathrm{G}$ sweeps~\cite{Dorow2022, Provias2023}, see Figure~\ref{fig:perfeval}(c) and Section~SI\ref{subsec:SI:meashystBTI}.
In order to compare the hysteresis widths $\Delta V_\mathrm{H}$ or the BTI shifts $\Delta V_\mathrm{th}$ of different technologies, the response must be normalized by EOT~\cite{Knobloch2023a, Karl2025}, using experiments conducted at comparable electric fields, similar time scales and temperatures. 

\subsection{Dielectric Breakdown}
\label{subsec:TDDB}
When subjecting the gate insulator to high fields over extended periods, dielectric breakdown~(BD) is observed, which is caused by the sustained damage of charge carriers flowing through the insulator.
As BD is a dynamic process that proceeds in several stages, it is termed time-dependent dielectric breakdown~(TDDB). 
Initially, at a fixed $V_\mathrm{G}$ charges tunnel through the gate insulator, possibly through a TAT mechanism.
During operation more charge traps will form, increasing the leakage currents, referred to as stress-induced leakage currents~(SILC)~\cite{Ji2016, Luo2023}. Next, small discontinuities in $I_\mathrm{G}$ appear, marking the first soft BD events. During the wear-out phase, $I_\mathrm{G}$ gradually rises until a rapid jump in $I_\mathrm{G}$ indicates hard BD~\cite{Ranjan2018}. 
TDDB depends on the gate area, insulator thickness, dielectric constant, insulator morphology and thermal conductivity, see Figure~\ref{fig:perfeval}(b) and Section~SI\ref{subsec:SI:breakdownmeas}.
While most of the literature on 2D materials references a dielectric strength in $[\mathrm{MVcm^{-1}}]$, such values are misleading, as 
dielectrics will break after a sustained flow of carriers over a certain amount of time. 
Instead, TDDB should be analyzed on multiple MIS capacitors of varying areas and insulator thicknesses, evaluating both the ramp-rate dependent breakdown voltage~($V_\mathrm{BD}$) and its Weibull slope~($\beta_\mathrm{V_{BD}}$), see Figure~\ref{fig:perfeval}(c).
In layered dielectrics like hBN, a layer-by-layer breakdown mechanism has been observed~\cite{Hattori2015}, potentially involving the formation of defective conducting bridges between layers~\cite{Ducry2022}.
Furthermore, metals with a high cohesive energy, seem to slow down BD in hBN~\cite{Shen2024}. Consequently, BD should be studied on the complete MIS gate stack intended for the 2D FETs, as both the metal gate and the 2D semiconductor impact TDDB.

\subsection{Self-Heating}
\label{subsec:heat}
Due to Joule self-heating~(SH) during operation, both the device mobility and reliability can degrade.
In stacked NS FETs, all channels contribute to SH~\cite{Pop2006}, while the numerous interfaces act as thermal bottlenecks. 
To reduce SH, the thermal conductivity~($\mathcal{K}$) of the various device materials and the thermal boundary conductance~(TBC) of their interfaces should be high.
However, 2D materials have relatively poor TBC due to the vdW gap at their interfaces~\cite{Yalon2017, Kim2021}.
Figure~\ref{fig:perfeval}(f) illustrates the expected TBC and in-plane $\mathcal{K}$ of monolayer MoS$_2$ interfaced with several materials; a TBC around $\SI{20}{MWm^{-2}K^{-1}}$ is relatively low~\cite{Yalon2017} and $\mathcal{K}$ depends on the insulator~\cite{Gabourie2022}, potentially being lower when MoS$_2$ is fully encased~\cite{Gabourie2021}, as in 2D NS FETs.
``Thermally-short'' (sub-\SI{150}{nm}) 2D FETs appear to dissipate heat mostly via their contacts~\cite{Gabourie2022}. 
In channels shorter than 10-\SI{15}{nm}, much of the heat will be generated at the contacts, due to quasi-ballistic transport in the channel~\cite{Pop2006}. 
Thus, the \textit{thermal resistance} of 2D contacts will become increasingly important, alongside their electrical resistance. 
Experimentally, Raman thermometry on micron-scale devices~\cite{Yalon2017, Vaziri2019} has been used to extract some thermal properties of interest. 
However, future work must evaluate and improve the TBC with the contacts, while nanoscale and GAA devices will need to rely on comprehensive modeling efforts~\cite{Gabourie2024} where measurements cannot be directly applied.

\section{Dielectric Candidates}
\label{sec:insulators}
\begin{figure}[!ht]
\centering
\includegraphics[width=\linewidth]{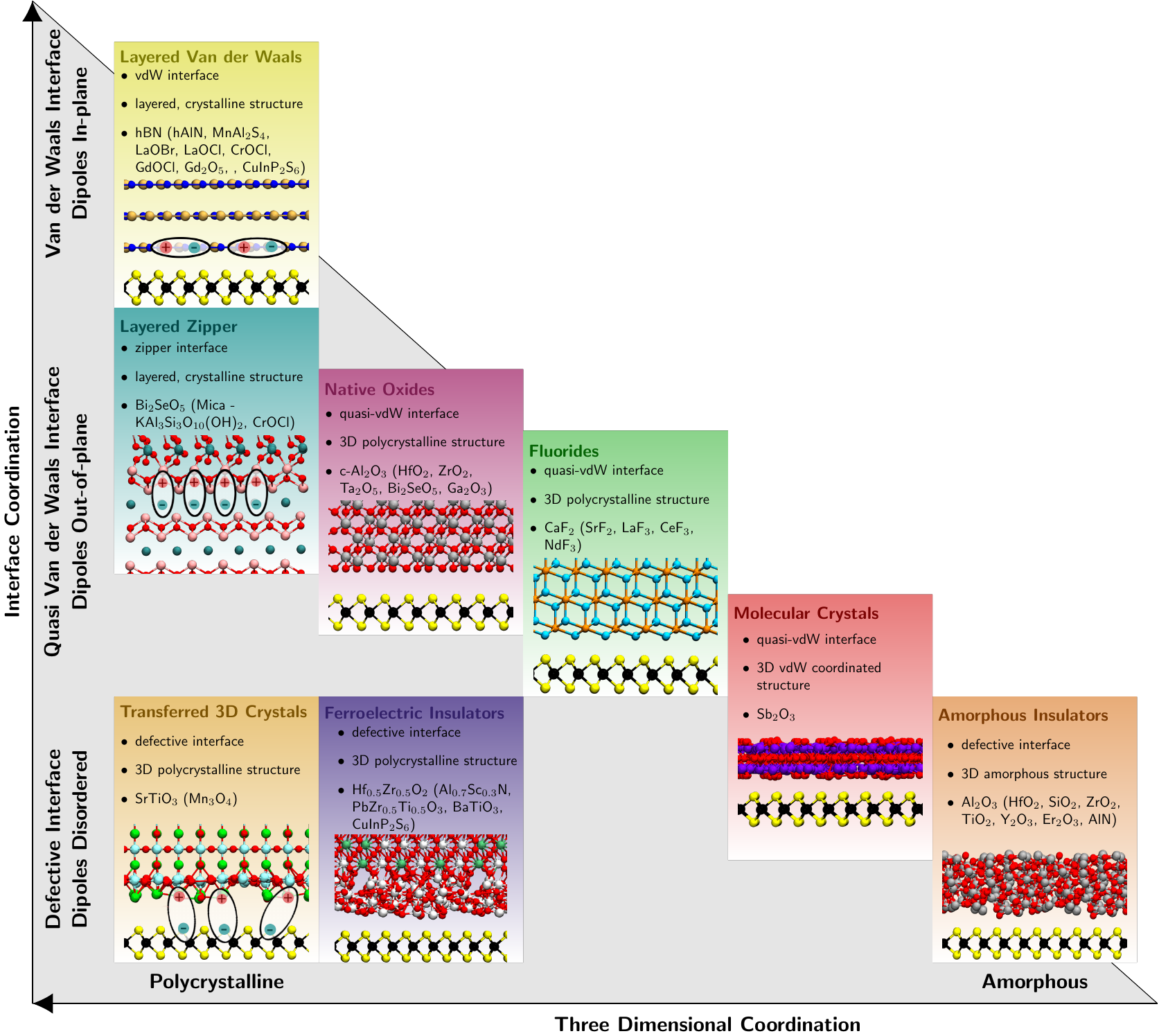}
\caption{Materials that could serve as gate dielectrics for 2D FETs, classified by their interface formed with 2D layered semiconductors and their crystalline structure (N-blue, B-ocher, S-yellow, Mo-black, O-red, Se-turquoise, Bi-pink, Al-dark gray, Ca-orange, F-cyan, Sr-light green, Ti-light blue, Zr-dark green, Hf-light gray, Sb-purple). Dielectrics are listed multiple times if they are part of different groups. }
\label{fig:insulators}
\end{figure}

To date, over 30 dielectrics in various phases have been proposed and experimentally realized for use in 2D FETs.  
These dielectrics can be classified by two main criteria, first, their material structure is either amorphous or polycrystalline.
Second, the interface formed between a 2D material and the gate insulator is either a van der Waals~(vdW) interface, a quasi-vdW interface~\cite{Koma1992} or a defective interface, depending on the interface quality and the dipole alignment~\cite{Woomer2019}.
Interfacial dipoles can be in-plane, out-of-plane or in a disordered way and impact phonon scattering, see Subsection~\ref{sec:performance}\ref{subsec:mobility}.
With these two criteria, we can create a visual summary of all potential gate dielectrics using eight major groups, see Figure~\ref{fig:insulators}.
These groups are layered vdW dielectrics~\cite{Britnell2012a, Osanloo2021} (Subsection~\ref{subsec:vdW}), layered zipper dielectrics~\cite{Li2020c, Tang2025}, (Subsection~\ref{subsec:zipper}), native oxides~\cite{Kang2024,Zeng2024}(Subsection~\ref{subsec:natox}), fluorides~\cite{Illarionov2019, Meng2024}(Subsection~\ref{subsec:fluorides}), transferred 3D crystals~\cite{Yang2022, Huang2022} (Subsection~\ref{subsec:transcrystals}), ferroelectric dielectrics~\cite{Ko2016, Si2018} (Subsection~\ref{subsec:ferro}), inorganic molecular crystals~\cite{Liu2021b, Xu2023} (Subsection~\ref{subsec:molecular}) and amorphous dielectrics~\cite{Li2019, Mortelmans2024} (Subsection~\ref{subsec:amorphous}). 
Gate dielectrics that have so far been used in 2D FET prototypes are summarized in Table~\ref{tab:insulators} and in the following the dielectrics, reported prototypes and their respective potential are discussed.

\subsection{Layered Van der Waals Dielectrics}
\label{subsec:vdW}
Hexagonal boron nitride~(hBN) is the most widely used layered vdW insulator~\cite{Lan2024, Shen2025}. 
It provides a high quality interface and low scattering in adjacent 2D semiconductors~\cite{Liu2023}, but has a small dielectric constant of  $\sim 3.8$~\cite{Laturia2018}, causing high gate leakage at a CET of 
\SI{0.9}{nm}~\cite{Britnell2012a, Knobloch2021}, see Subsection~\ref{sec:performance}\ref{subsec:leakage} and Subsection~SI\ref{subsec:SI:hBNleakage}. 
This can be mitigated by combining monolayer hBN with a high-$\kappa$ gate dielectric, although hBN transfer is required~\cite{Lan2024, Wang2024}. 
In contrast, hexagonal aluminum nitride~(hAlN) can be epitaxially grown with PEALD on TMDs at \SI{250}{\degree C} and provides a dielectric constant of about 8.7~\cite{Chang2022, Wang2023}.
Besides hBN and hAlN, there are many ternary layered vdW dielectrics, including manganese aluminum sulfide~(MnAl$_2$S$_4$)~\cite{Xu2022}, and transition metal nitride halides~\cite{Osanloo2021}, such as lanthanum oxybromide~(LaOBr)~\cite{ Soll2024}, lanthanum oxychloride~(LaOCl)~\cite{Zhang2023}, chromium oxychloride~(CrOCl)~\cite{Guo2024}, or gadolinium oxychloride~(GdOCl)~\cite{Xu2024}, that typically require transfer.
Another layered vdW insulator is gadolinium pentoxide~(Gd$_2$O$_5$)~\cite{Yin2025}, the only layered vdW insulator so far for which an EOT below \SI{2}{nm} has been demonstrated~\cite{Yin2025}. 

\subsection{Layered Zipper Dielectrics}
\label{subsec:zipper}
Layered zipper materials have no vdW gap because of the stronger interlayer bonding in comparison to layered materials,
yet the interlayer bond strength remains smaller than typical covalent bonding.
They are characterized by an out-of-plane dipole moment and interfaces formed between adjacent layers where an ionic species covers 50\% of every surface, for example, Se atoms in bismuth oxyselenide, Bi$_2$O$_2$Se~\cite{Wei2019}.
Bi$_2$O$_2$Se can be oxidized into its native oxide bismuth oxoselenate, Bi$_2$SeO$_5$ in a layer-by-layer fashion in a UV-assisted oxidation process at about \SI{100}{\degree C}~\cite{Zhang2022, Tan2023, Tang2025}, see Section~\ref{sec:fabrication}\ref{subsec:nativeox}.
As Bi$_2$SeO$_5$ can conformally coat any geometry, one can fabricate FinFETs based on Bi$_2$O$_2$Se fins~\cite{Tan2023} or NS FETs with high on-currents of $\SI{0.3}{mA}$\textmu$\mathrm{m^{-1}}$ at IRDS conditions~\cite{Tang2025}. 
Another layered zipper insulator is mica~(KAl$_3$Si$_3$O$_{10}$(OH)$_2$), where adjacent layers are terminated by potassium interlayers with 50\% coverage~\cite{Franceschi2023}.
Mica is widely available and has been exfoliated down to thicknesses of about \SI{10}{nm}, even though the reliance on mechanical exfoliation has so far led to poor reproducibility and large variability in mica-based FETs~\cite{Zou2019, Zou2020}.

\subsection{Native Oxides}
\label{subsec:natox}
Native oxides can be grown conformally and provide a high quality interface, making them promising for stacked 2D NS FETs, even though it is difficult to selectively oxidize the outer layers without damaging the channel.
Besides Bi$_2$SeO$_5$ as oxide to Bi$_2$O$_2$Se~\cite{Li2020c, Tang2025}, 
HfS$_2$ or HfSe$_2$ can be oxidized into HfO$_2$~\cite{Mleczko2017, Kang2024}, ZrS$_2$ or ZrSe$_2$ into ZrO$_2$~\cite{Jin2023}, and TaS$_2$ into Ta$_2$O$_5$~\cite{Chamlagain2017a}. 
Unfortunately, MoO$_3$, the oxide of MoS$_2$, has a large electron affinity and forms a staggered band gap~\cite{Guo2014b}. Hence, MoO$_\mathrm{x}$ cannot be used as a gate oxide on its own, even though controlled oxidation is possible~\cite{Reidy2023}. 
Nevertheless, both WO$_\mathrm{x}$~(oxide of WSe$_2$) and MoO$_\mathrm{x}$ can serve as seeding materials for ALD of HfO$_2$ and as interlayers in gate stacks that p-dope WSe$_2$ channels~\cite{Pang2020a}. 
In addition, a metal oxide can also serve as gate oxide, for example amorphous Al$_2$O$_3$~\cite{English2017}, gallium oxide~(Ga$_2$O$_3$)~\cite{Yi2024} or crystalline aluminum oxide~(c-Al$_2$O$_3$)~\cite{Zeng2024}, even though current prototypes of Ga$_2$O$_3$ and c-Al$_2$O$_3$ involved an oxide transfer~\cite{Yi2024, Zeng2024}.

\subsection{Fluorides}
\label{subsec:fluorides}
Since fluorides are typically grown using MBE on lattice-matched substrates, they are limited to planar device geometries~\cite{Illarionov2019, Wen2020}.
Back-gated MoS$_2$ FETs with a gate dielectric of \SI{2}{nm} calcium fluoride~(CaF$_2$) have been demonstrated with good stability~\cite{Illarionov2019, Illarionov2019a}.
Yet, the fluorine-termination of the surface causes strong phonon coupling, thereby limiting the MoS$_2$ mobility, see Subsection~\ref{sec:performance}\ref{subsec:mobility}. 
Recently, amorphous lanthanum trifluoride~(LaF$_3$) deposited at room temperature was used as a back-gate insulator for n-type MoS$_2$ and p-type WSe$_2$ FETs~\cite{Meng2024}.
In superionic LaF$_3$, the F$^-$ ions can move around due to large concentrations of vacancies and interstitials, contrary to MBE CaF$_2$ films.
These F$^-$ ions accumulate at the interface and form an electric double layer providing high effective capacitances, however, the capacitance degrades at frequencies above \SI{100}{Hz}, rendering superionic fluorides unsuitable for digital logic~\cite{Meng2024}.

\subsection{Transferred 3D Crystals}
\label{subsec:transcrystals}
Three-dimensional~(3D) crystals, for example perovskites, can be transferred onto 2D semiconductors, using a sacrificial oxide in the release process, see Subsection~SI\ref{subsec:SI:transfer}. 
Perovskites show promise as gate dielectrics in scaled FETs because of their high dielectric constants~\cite{Yang2023}, theoretically reaching over 300 in strontium titanate~(SrTiO$_3$, STO).
However, permittivity depends heavily on the layer thickness due to dead layer effects, which reduce the permittivity down to about 30 in thin layers~\cite{Huang2022, Huang2022a}, see Subsection~\ref{sec:performance}\ref{subsec:capacitance}. 
By transferring thin STO membranes, an EOT below \SI{2}{nm} is possible~\cite{Huang2022, Huang2022a}, see Table~\ref{tab:insulators}.
Another 3D crystal with a high permittivity of over 80 in thin films that can be transferred to form top-gated FETs is manganese oxide~(Mn$_3$O$_4$)~\cite{Yuan2025}. 
The main drawback of the transfer process is that it is not scalable, see Subsection~\ref{sec:fabrication}\ref{subsec:transfer}.

\subsection{Ferroelectric Insulators}
\label{subsec:ferro}
Ferroelectric gate insulators enable threshold voltage switching via their remanent polarization, supporting fast low-power non-volatile FeFET memories~\cite{Ko2016, Liu2021a}.
This allows both reconfigurable digital~\cite{Wu2020} and analog~\cite{Zhu2023} electronics, including neuromorphic circuits~\cite{Shen2020}.
Ferroelectrics have also inspired Negative Capacitance FETs~(NCFETs), targeting sub-\SI{60}{mV/dec.} subthreshold swings to surpass the thermionic limit, though their viability is debated~\cite{Cao2020}.
Hafnium zirconium oxide~(HZO), grown by ALD, is currently the leading material option thanks to CMOS compatibility and robust ferroelectricity at scaled thicknesses~\cite{Mcguire2017, Shen2020}. 
These are properties that ferroelectric perovskites, such as barium titanate~(BaTiO$_3$, BTO)~\cite{Puebla2022} or lead zirconium titanate~(PbZr$_{0.5}$Ti$_{0.5}$O$_{3}$, PZT)~\cite{Ko2016} lack. 
Alternatively, AlScN offers high polarization and CMOS-compatible sputtering~\cite{Liu2021a}, but conformal deposition remains a challenge.
Recently, vdW ferroelectrics such as CuInP$_2$S$_6$~(CIPS), with a band gap of about \SI{2.7}{eV}~\cite{Si2018a}, have been integrated as a gate dielectric in MoS$_2$~\cite{Si2018a} and WS$_2$~\cite{Zhao2023} FETs, expanding the options for 2D ferroelectric devices.

\subsection{Inorganic Molecular Crystals}
\label{subsec:molecular}
Inorganic crystals are characterized by a 3D vdW-coordinated structure.
For instance, Sb$_2$O$_3$ molecules form bicyclic cages with loose intermolecular vdW bonds, enabling high-quality vdW interfaces with 2D semiconductors, yet leading to
fast dielectric breakdown~\cite{Liu2021b}, see Subsection~\ref{sec:performance}\ref{subsec:TDDB}.
Sb$_2$O$_3$ can be grown using CVD on mica~\cite{Han2019} or it can be thermally evaporated on silicon, SiO$_2$~\cite{Liu2021b}, or 2D semiconductors like MoS$_2$~\cite{Xu2023}.  Sb$_2$O$_3$ can serve as a seeding layer for ALD of HfO$_2$ on TMDs~\cite{Xu2023}, achieving a small EOT of \SI{0.7}{nm}, see Table~\ref{tab:insulators}, even though large-scale uniformity remains challenging.

\subsection{Amorphous Dielectrics}
\label{subsec:amorphous}
Due to their use as high-$\kappa$ gate dielectrics in silicon technologies, amorphous dielectrics are the most studied insulator category for 2D devices. 
Amorphous dielectrics already used in 2D FETs include Al$_2$O$_3$~\cite{Park2016f, Wirtz2015, Lin2020}, HfO$_2$~\cite{Jeong2016a, Price2019, Li2019, Wu2021a, Luo2022, Mortelmans2024}, SiO$_2$~\cite{Smithe2017a, Ho2023}, ZrO$_2$~\cite{Zhao2022, Jin2023}, titanium dioxide~(TiO$_2$)~\cite{Kropp2018}, yttrium oxide~(Y$_2$O$_3$)~\cite{Zou2014, Wang2022}, erbium oxide~(Er$_2$O$_3$)~\cite{Uchiyama2023}, tantalum oxide~(Ta$_2$O$_5$)~\cite{Chamlagain2017a, Lan2023}, aluminum nitride~(AlN)~\cite{Liu2021a}, and silicon nitride~(Si$_3$N$_4$)~\cite{ Ho2023}. 
Typically, these dielectrics are deposited via ALD~\cite{Zou2014, Price2019}, see Subsection~\ref{sec:fabrication}\ref{subsec:ALD}, generally requiring a seed layer for deposition on top of 2D semiconductors~\cite{Park2016f, Li2019}.
ALD Al$_2$O$_3$/HfO$_2$ gate stacks are the currently preferred method for 2D devices fabricated in industry-compatible environments~\cite{Lin2020, Smets2021a, OBrien2021, Chung2022, Dorow2022, Chou2023, Chung2024}. 
From these industry-compatible gate stacks, so far only a few have achieved a scaled EOT below 2\,nm ~\cite{Dorow2022, Chou2023, Chung2024, Mortelmans2024}, while some university laboratories have reported amorphous dielectrics with sub-1\,nm EOT~\cite{Wang2023, Xu2023, Uchiyama2023}, even though using evaporated seed layers renders them likely unsuitable for conformally coating GAA devices. 
Moreover, the rigorous evaluation of CET through $C_\mathrm{G}(V_\mathrm{G})$ measurements is lacking for most demonstrations, see Section~\ref{sec:performance}\ref{subsec:capacitance}.

\section{Performance Potential of Dielectrics}
\label{sec:perfpotential}
To assess the performance potential of dielectrics, we modeled gate leakage using the Tsu-Esaki model and used the Boltzmann Transport Equation~(BTE) to assess mobility in an adjacent MoS$_2$ channel. 
First, we compared leakage currents at an EOT of 1\,nm, see Section~\ref{sec:performance}\ref{subsec:leakage} and Section~SI\ref{subsec:SI:method_leak}.
Figure~\ref{fig:perfcalc}(a) shows the calculated leakage current density as a function of gate bias.
For p-FETs, a WSe$_2$ monolayer channel was used, while MoS$_2$ served as channel material for n-type FETs,  with the exception of Bi$_2$O$_2$Se that was paired with its native oxide Bi$_2$SeO$_5$. The gate contact is gold, see the band alignment in Figure~\ref{fig:perfcalc}(b).
Figures~\ref{fig:perfcalc}(a) and~\ref{fig:perfcalc}(c) reveal that hBN and SiO$_2$ cannot sufficiently block leakage currents at an ultimately scaled EOT due to their small dielectric constants. Even in the defect-free scenario, the tunnel currents through hBN and SiO$_2$ exceed the low-power limit of 
\SI{700}{mAcm^{-2}}~\cite{IRDS2023}, see Table~\ref{tab:IRDS}.
In contrast, SrTiO$_3$, Al$_2$O$_3$, CaF$_2$, HfO$_2$, and Bi$_2$SeO$_5$  sufficiently block currents at an EOT of \SI{0.5}{nm} and a $V_\mathrm{DD}$ of \SI{0.6}{V}.
The estimated tunnel currents also depend on the $\sim\SI{0.3}{nm}$ vdW gap between the dielectric and the 2D semiconductor.
Table~\ref{tab:materials} summarizes relevant material parameters of the analyzed dielectrics.
Figure~\ref{fig:perfcalc}(c) compares calculated leakage currents with measured current densities reported in literature.
The measured leakage currents are higher than the theoretical predictions due to the presence of defects.
\textit{At the same time, correctly determining the EOT of the measured gate dielectrics is challenging, as CET is the actual measurement quantity} (see Subsection~\ref{sec:performance}\ref{subsec:capacitance}) and it is often difficult to reliably rule out the presence of a thin layer of organic contamination at the 2D/insulator interface~\cite{Tilmann2023} that might reduce leakage currents, see Section~SI~\ref{subsec:SI:hBNleakage}.
To quantify how surrounding insulators affect mobility, we solved the BTE for monolayer MoS$_2$ sandwiched between two insulators in a double-gate configuration, see Subsection~\ref{sec:performance}\ref{subsec:mobility} and Section~SI~\ref{subsec:SI:method_mobility}. 
These values represent upper mobility limits, refined calculations may include the impact of long-range electrostatics on the 2D material phonons~\cite{Sohier2017, Royo2021} or the impact of potential fluctuations on the order of \SI{200}{meV} at the atomic scale due to the defective interface between 2D semiconductors and ALD oxides~\cite{Vantroeye2025}.
Figure~\ref{fig:perfcalc}(d) plots the mobility as a function of the dielectric constant and the charged impurity concentration.
As charged impurities are screened more in dielectrics with high permittivity, their impact is largest at interfaces with dielectrics with small dielectric constants like hBN. The IRDS goal of $\SI{60}{cm^2V^{-1}s^{-1}}$~\cite{IRDS2023}, see Table~\ref{tab:IRDS}, seems attainable in monolayer MoS$_2$, but most likely not if MoS$_2$ is surrounded by CaF$_2$, HfO$_2$ or ZrO$_2$. 
Instead, an hAlN or Al$_2$O$_3$ interlayer in combination with HfO$_2$~\cite{Wang2023, Chung2024}, a complete Al$_2$O$_3$ gate stack~\cite{Zeng2024}, or a Bi$_2$SeO$_5$ native oxide to Bi$_2$O$_2$Se channels~\cite{Tang2025} are promising candidates for facilitating high mobilities.
Finally, we compare the experimental performance of the most promising 2D FET prototypes in Figures~\ref{fig:perfcalc}(e)-(g) and in Table~\ref{tab:insulators} at IRDS conditions for the so-called $\SI{5}{\AA}$ high density node, see Table~\ref{tab:IRDS}~\cite{IRDS2023}. 
So far, no studies have analyzed the proposed gate stacks at exactly these conditions, hence we used conditions that were as close as possible.
On-currents are typically higher in 2D prototypes from industry due to better 2D film quality and smaller contact resistances.
At the same time, the defective interface between ALD HfO$_2$ and the 2D channel makes it difficult to reach a scaled EOT at a small $SS_\mathrm{avg}$, which is often achieved for native oxides and transferred dielectrics in academia. 
However, in these comparisons the gate length is not specified, even though $SS_\mathrm{avg}$ will be limited by short-channel effects, hence future studies must focus on evaluating scaled devices with $L_\mathrm{G}<\SI{20}{nm}$.

\section{Conclusions and Outlook}
The primary goal for the gate stack is to provide excellent gate control, requiring a large gate capacitance, hence $\mathrm{CET}<\SI{0.9}{nm}$ and $t_\mathrm{ins}<\SI{3}{nm}$, thereby suppressing short-channel effects and providing steep switching with $SS_\mathrm{avg}<\SI{65}{mV/dec.}$ at scaled gate lengths $L_\mathrm{G}<\SI{12}{nm}$. At the same time, the interface of the 2D TMD with the insulator must not degrade the semiconductor mobility via remote phonon and impurity scattering, enabling on-current densities above $\SI{600}{\micro\ampere\per\micro\meter}$.
These goals need to be reached when using conformal deposition methods that allow integration into a gate-last fabrication of stacked 2D NS FETs, most likely ALD or oxidation. 
Currently, promising insulator candidates include the native oxides Bi$_2$SeO$_5$~\cite{Tang2025} or HfO$_2$~\cite{Kang2024}, or ALD layers, for example HfO$_2$~\cite{Mortelmans2024} or Er$_2$O$_3$~\cite{Uchiyama2023}, potentially with interlayers like hAlN~\cite{Wang2023}, Al$_2$O$_3$~\cite{Jiang2024}, SiO$_2$~\cite{Ko2025a}, GdAlO$_x$~\cite{Wu2021a}, Sb$_2$O$_3$~\cite{Xu2023} or Ta$_2$O$_5$~\cite{Lan2023}. 
To realistically assess insulator performance, prototype devices employing these various gate stacks should be fabricated with a CET below \SI{0.9}{nm}, as evaluated by $C_\mathrm{G}\left(V_\mathrm{G}\right)$ measurements. 
Moreover, the main performance metrics should be extracted at the operation conditions stated in the IRDS for the 2037 ``\SI{5}{\AA}'' node~\cite{IRDS2023}, supported by sufficient statistics to capture device-to-device variability. 
In general, any gate stack for 2D NS FETs needs to provide small variability and high reliability, which has not been achieved so far.
Interface trap densities of 2D devices remain high, necessitating a more careful analysis of the origins of large interface trap densities based on $C_\mathrm{G}\left(V_\mathrm{G}\right)$ measurements and admittance spectroscopy. 
Strategies to define $V_\mathrm{th}$ in 2D FETs for stable enhancement-mode operation with minimal variability have also remained elusive.
Furthermore, reliability tests need to address bias temperature instabilities for both n- and p-type devices,
 across small insulator thicknesses, at elevated temperatures, and under a set of stress biases.
Additionally, dielectric breakdown studies with sufficient statistics for the most relevant MIS gate stacks are needed.
Finally, the thermal boundary conductances of promising gate dielectrics for 2D FETs should be evaluated, because for stacked channels, self-heating might become a critical issue, even though short-channel ballistic transport may mitigate some heating.

In summary, the progress reported for 2D FETs in the last decade has been impressive. 
The fabrication and high-performance operation of nanoscaled stacked 2D FETs seems to be within reach, but several critical challenges related to the gate dielectrics need to be addressed before this novel technology can be considered for commercial applications.
Future studies must focus on reporting the performance of scaled 2D FETs as closely as possible to the most realistic BEOL or FEOL operating conditions, while also devoting considerably more efforts to studying variability and reliability in the gate stacks with the greatest potential for scaled 2D FET operation. 
By collaboratively addressing these issues, we believe that the research community can make scaled, energy-efficient 2D FETs a commercially successful technology.




\section*{References}

\bibliography{refs_new}


\section*{Acknowledgments}
The authors wish to thank Benoit van Troeye, Ben Kaczer, Ernest Wu, Inge Asselberghs, and Mahdi Pourfath for valuable discussions.

\newpage
\normalsize

\section*{Figures}


\newpage

\begin{figure}[!ht]
\centering
\includegraphics[width=.9\linewidth]{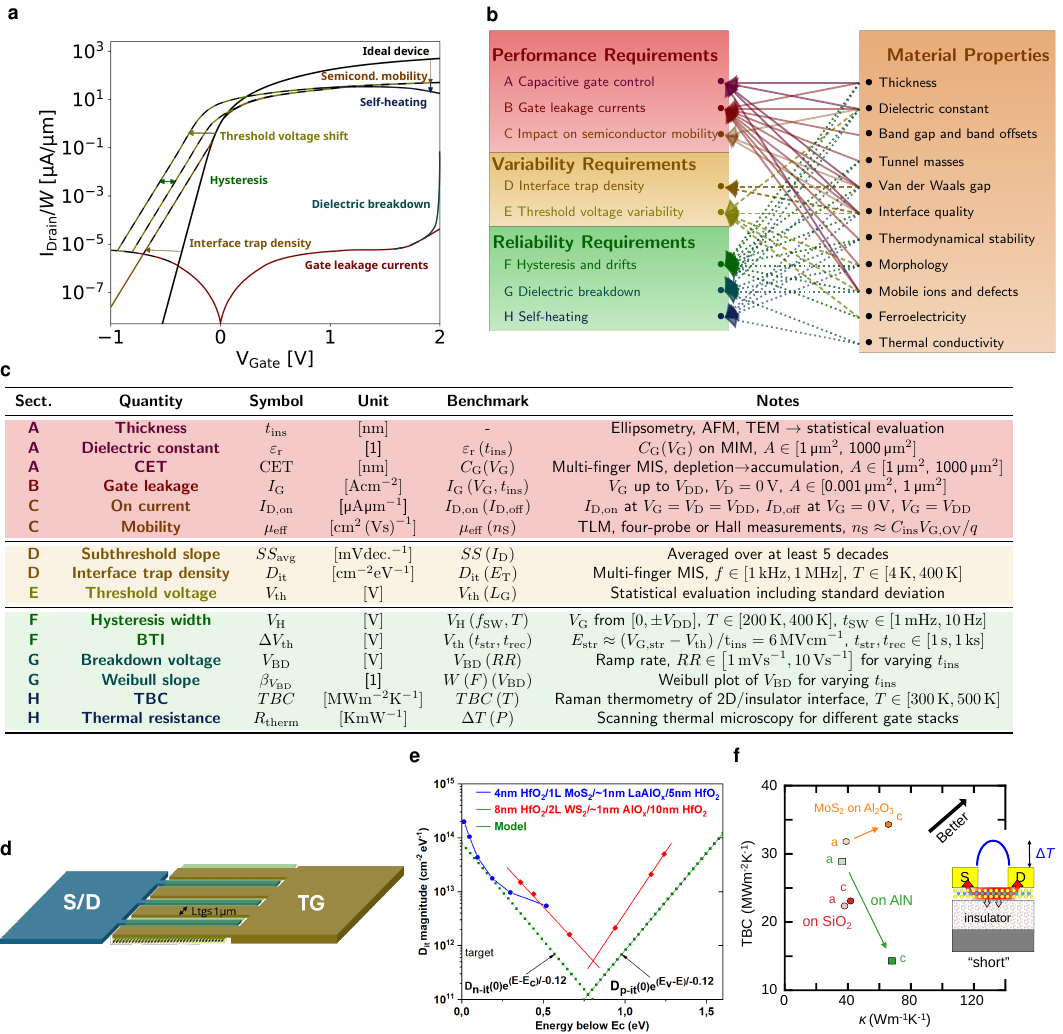}
\caption{Performance and reliability requirements of gate dielectrics: 
\textbf{(a)} Overview about how the requirements for suitable gate dielectrics for 2D FETs affect the transfer characteristics and hence the performance.
\textbf{(b)} Schematic listing the requirements for good gate dielectrics for 2D FETs and how they depend on the material properties of the dielectric.
\textbf{(c)} Suggested benchmarking quantities for evaluating the gate-insulator related performance, variability and reliability of the FET. Here, $E_\mathrm{T}$ is the interface trap level, hence the energetic position of the interface trap in $\mathrm{eV}$, $W(F)(V_\mathrm{BD})$ is the Weibit function, the Weibull cumulative distribution function of the breakdown voltage $V_\mathrm{BD}$, and $P$ is the heating power in $\mathrm{W}$. 
\textbf{(d)} Multi-finger capacitor structure for admittance measurements of 2D gate stacks. The charge is injected laterally from the S/D fingers. $L_\mathrm{tg}$ is kept sufficiently short $<\SI{1}{}$\textmu$\mathrm{m}$ to lower the series resistance, while the multi-finger design boosts the total area enabling admittance measurements.
\textbf{(e)} $D_\mathrm{it}\left(E\right)$ profiles are obtained using multi-finger structures with multi-frequency $C_\mathrm{G}\left(V_\mathrm{G}\right)$. Lower temperatures enable probing close to the band edges, where exponentially increasing $D_\mathrm{it}$ is extracted for both gate stacks. Admittance measurements on a gate stack of TiN/4nm HfO$_2$/1L MoS$_2$/1nm LaAlO$_\mathrm{x}$/5nm HfO$_2$ and on a gate stack of TiN/8nm HfO$_2$/1L MoS$_2$/1nm AlO$_\mathrm{x}$/8nm HfO$_2$, adapted from~\cite{Mootheri2021}. 
\textbf{(f)} Thermal boundary conductance (TBC) vs. thermal conductivity ($\mathcal{K}$) for a common 2D semiconductor (monolayer MoS$_2$) interfaced with various amorphous (a) and crystalline (c) insulators~\cite{Gabourie2022}. A trade-off may exist depending on the phonon mode overlap between the 2D material and insulator. Inset schematic shows temperature rise ($\Delta T$) and heat flow paths in a ``thermally-short'' device ($< \SI{150}{nm}$), wherein more heat is dissipated to the source/drain (S/D) contacts than to the insulator~\cite{Gabourie2021}.
}
\label{fig:perfeval}
\end{figure}

\newpage

\newpage

\begin{figure}[!ht]
\includegraphics[width=\linewidth]{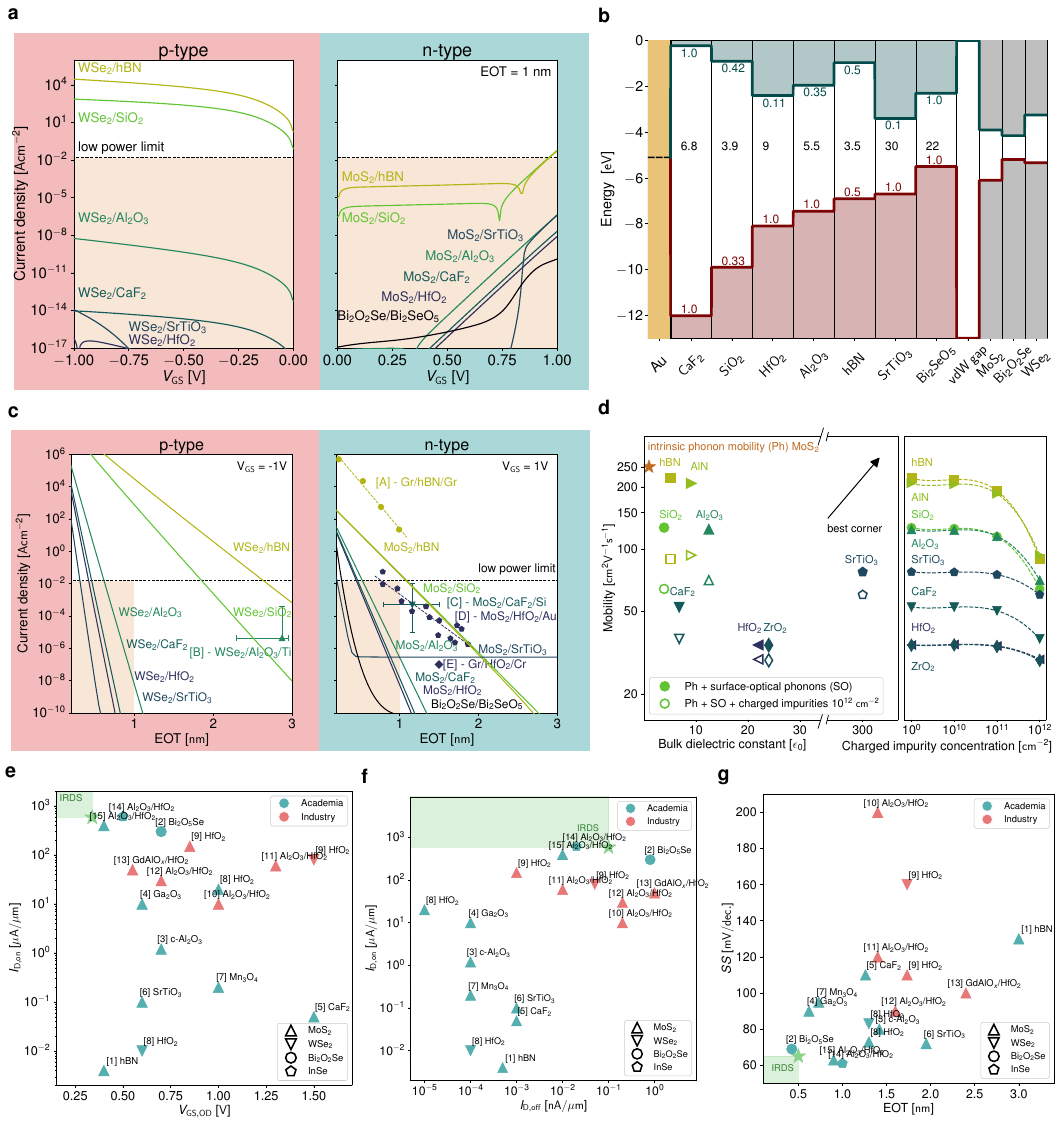}
\caption{Performance potential of gate dielectrics: \textbf{(a)} Comparison of the gate leakage currents for p-type and n-type FETs at a constant EOT of \SI{1}{nm} based on a Tsu-Esaki approximation of the tunnel current through a defect-free insulator using Au as gate metal and \SI{0.3}{nm} vdW gap except for Bi$_2$SeO$_5$ with no vdW gap.  \textbf{(b)} Band diagram of gate dielectrics where leakage currents are compared with the effective tunnel masses for electrons and holes~($m_\mathrm{tun,e}$, $m_\mathrm{tun,h}$) given at the band edges and typical reported values of the dielectric constant for thin layers~($\varepsilon_\mathrm{r}$) in the middle. The literature references are given in Table~\ref{tab:materials}. 
\textbf{(c)} Comparison of the leakage currents for p-type and n-type FETs at a constant gate bias as a function of EOT with the orange shaded corner highlighting the target region, including measured leakage currents: [A]~\cite{Britnell2012a}, [B]~\cite{Park2016f}, [C]~\cite{Illarionov2019}, [D]~\cite{Li2019}, [E]~\cite{Jeong2016a}.
\textbf{(d)} Comparison of the mobility in monolayer MoS$_2$ including the contributions of surface optical phonons (SO) and interfacial charged impurities of varying concentrations for different dielectric substrates.
\textbf{(e)-(g)} Performance comparison of recent 2D FET prototypes with the performance requirements as stated in the IRDS at room temperature~\cite{IRDS2023}. The numbers correspond to the following references: [1]~\cite{Shen2025}, [2]~\cite{Tang2025}, [3]~\cite{Zeng2024}, [4]~\cite{Yi2024}, [5]~\cite{Illarionov2019}, [6]~\cite{Huang2022}, [7]~\cite{Yuan2025}, [8]~\cite{Li2019}, [9]~\cite{Mortelmans2024}, [10]~\cite{Chung2024}, [11]~\cite{Chou2023}, [12]~\cite{Dorow2022}, [13]~\cite{Wu2021a}. Here $V_\mathrm{GS,OD}$ is the gate overdrive voltage where the on-current was reported, i.e., $V_\mathrm{GS,OD} = V_\mathrm{G}-V_\mathrm{th}$.
}
\label{fig:perfcalc}
\end{figure}

\newpage


\section*{Supplementary Information}

\begin{table}[!ht]
\begin{tabular}{ccccccccc}
\toprule
Quantity & Symbol & Unit & \multicolumn{6}{c}{Node} \\
\midrule
 Year of production & &  & 2025 & 2025 & 2031& 2031& 2037&2037\\
 Node name & & & ``\SI{2}{nm}'' & ``\SI{2}{nm}'' & ``\SI{10}{\AA}'' & ``\SI{10}{\AA}'' & ``\SI{5}{\AA}'' & ``\SI{5}{\AA}''\\
 Target application & & & High Power & High Density & High Power & High Density & High Power & High Density \\
\midrule
Device design & & & GAA & GAA & CFET & CFET & CFET & CFET \\
Gate length & $L_\mathrm{G}$ & $[\mathrm{nm}]$ & 14 & 14 & 12 & 12 & 12 & 12\\
Number of sheets & & & 3 & 3 & 5 & 5 & 5 & 5\\
Width per sheet& $W$ & $[\mathrm{nm}]$ & 30 & 15 & 15 & 10 & 15 & 6\\
Effective total width &  $W_\mathrm{eff}$ & $[\mathrm{nm}]$ & 216 & 126 & 210 & 160 & 190 & 100\\
\midrule 
Supply voltage & $V_\mathrm{DD}$ &$[\mathrm{V}]$& 0.65 & 0.65 & 0.6 & 0.6 & 0.6 & 0.6\\
Subthreshold slope & $SS$ & $[\mathrm{mV/dec.}]$ & 72 & 67 & 70 & 65 & 70 & 65\\
Capacitance equivalent thickness & CET & $[\mathrm{nm}]$ & 1.0 & 1.0 & 0.9 & 0.9 & 0.9 & 0.9 \\
Channel capacitance correction & $t_\mathrm{sc, eff}$ & $[\mathrm{nm}]$ & 0.1 & 0.1 & 0.1 & 0.1 & 0.1 & 0.1 \\
Van der Waals gap & $t_\mathrm{vdW}$ & $[\mathrm{nm}]$ & 0.3 & 0.3 & 0.3 & 0.3 & 0.3 & 0.3 \\
Equivalent oxide thickness & EOT & $[\mathrm{nm}]$ & 0.6 & 0.6 & 0.5 & 0.5 & 0.5 & 0.5 \\
Threshold voltage & $V_\mathrm{th}$ & $[\mathrm{V}]$ & 0.165 & 0.281 & 0.164 & 0.274 & 0.161 & 0.261\\
Overdrive voltage & $V_\mathrm{OV}$ & $[\mathrm{V}]$ &  0.485 & 0.369 & 0.436 & 0.326 & 0.439 & 0.339 \\
Off current & $I_\mathrm{off}$ & [$\mathrm{nA}$ \textmu $\mathrm{m^{-1}}$] & 10 & 0.1 & 10 & 0.1 & 10 & 0.1 \\ 
Off current density & $I_\mathrm{off}$ & $[\mathrm{A cm^{-2}}]$ & 70 & 0.7 & 80 & 0.8  & 80 & 0.8  \\
On current & $I_\mathrm{D,on}$ & [\textmu $\mathrm{A}$\textmu $\mathrm{m^{-1}}$] & 787 & 602 & 775 & 562& 790 & 587 \\
On/off ratio & $I_\mathrm{on}/I_\mathrm{off}$ & $[1]$ & \SI{e5}{} & \SI{e7}{} & \SI{e5}{} & \SI{e7}{} & \SI{e5}{} & \SI{e7}{} \\ 
Effective mobility & $\mu_\mathrm{eff}$ & $[\mathrm{cm^2 \left(Vs\right)^{-1}}]$ & 100 & 100 & 80 & 80 & 60 & 60 \\
Intrinsic delay & $\tau_\mathrm{delay}$ & $[\mathrm{ps}]$ & 1.06 & 1.06 & 0.86 & 0.86 & 0.84 & 0.84 \\
\bottomrule
\end{tabular}
\caption{IRDS requirements. The data for $L_
\mathrm{G}$, $W$, $W_\mathrm{eff}$, $V_\mathrm{DD}$, $SS$, $\mathsf{CET}$, $V_\mathrm{th}$, $I_\mathrm{D,on}$, $\mu_\mathrm{eff}$, and $\tau_\mathrm{delay}$ are taken directly from the current IRDS~\cite{IRDS2023}. The remaining quantities $t_\mathrm{sc,eff}$, $t_\mathrm{vdW}$, $\mathsf{EOT}$, $V_\mathrm{OV}$, $I_\mathrm{on}/I_\mathrm{off}$ are calculated based on the other quantities, assuming a 2D FET using a monolayer TMD as a channel material. }
\label{tab:IRDS}
\end{table}

\afterpage{
\clearpage
\thispagestyle{empty}
\begin{landscape}
\centering
\begin{tabular}{ccccccccccccc}
\toprule
\headcolor \textbf{Material} & $\varepsilon_\mathrm{r, \perp}$ bulk & $\varepsilon_\mathrm{r, \perp}$  at 0.5nm EOT & geom. & scal. & polar. & $t_\mathrm{ins}$ & EOT & $I_\mathrm{D,on}$ & $V_\mathrm{DD}$ & $SS$ & $V_\mathrm{th}$ & $I_\mathrm{G}$ at $V_\mathrm{DD}$\\
\headcolor - & $[1]$ & $[1]$ & - & - & - & $[\mathrm{nm}]$ & $[\mathrm{nm}]$ & [\textmu A\textmu$\mathrm{m}^{-1}$] & $[\mathrm{V}]$ & $[\mathrm{mV {dec.}^{-1}}]$ & $[\mathrm{V}]$ & $[\mathrm{mA{cm}^{-2}}]$\\
\midrule
\headcolor \multicolumn{4}{l}{\textbf{Layered Van der Waals Dielectrics}}
&\multicolumn{9}{c}{ } \\
hBN & 5.1~\cite{Geick1966} & 3.5~\cite{Laturia2018} & TG & no & n & 4~\cite{Shen2025} & 3~\cite{Shen2025} & 0.004~\cite{Shen2025} & 0.6~\cite{Shen2025} & 130~\cite{Shen2025} & +0.3(n)~\cite{Shen2025} & 0.5~\cite{Shen2025} \\
hAlN &  8.7~\cite{Chang2022} & - & - & -& -& -&-& -& -& -& -&-\\
MnAl$_2$S$_4$ & 6.1~\cite{Xu2022} & - & TG & no & n & 15~\cite{Xu2022} & 9.6~\cite{Xu2022} & 0.2~\cite{Xu2022} & 2~\cite{Xu2022} & 90~\cite{Xu2022} & -1.1(n)~\cite{Xu2022} & 0.01~\cite{Xu2022} \\
LaOBr & 9~\cite{Soll2024} & 13~\cite{Osanloo2021} & TG & no & n & 20~\cite{Soll2024} & 8.5~\cite{Soll2024} & 2~\cite{Soll2024} & 4~\cite{Soll2024}& 85~\cite{Soll2024} & -2(n)~\cite{Soll2024} & 0.02~\cite{Soll2024} \\
LaOCl &  11.8~\cite{Osanloo2021} & 50~\cite{Osanloo2021}&  loc. BG & no & n & 21~\cite{Zhang2023} & 7~\cite{Zhang2023} & 0.005~\cite{Zhang2023}& 6~\cite{Zhang2023} & 78~\cite{Zhang2023} & -1.5(n)~\cite{Zhang2023} & 1~\cite{Zhang2023}\\
CrOCl &  5~\cite{Guo2024} & -  & -& -& -& -& -& -& -& -&-&-\\
GdOCl & 15.3~\cite{Xu2024} & 15.3~\cite{Xu2024} & TG & no & n & 14.5~\cite{Xu2024} & 3.7~\cite{Xu2024} & 1.5~\cite{Xu2024} & 2~\cite{Xu2024}& 75~\cite{Xu2024} & -0.4(n)~\cite{Xu2024} & 0.003~\cite{Xu2024}\\
\rowyellow Gd$_2$O$_5$ & 25~\cite{Yin2025} & 19~\cite{Yin2025} & TG & no & n & 7.2~\cite{Yin2025} & 1.5~\cite{Yin2025} & 1~\cite{Yin2025} & 1~\cite{Yin2025} & 75~\cite{Yin2025} & -0.4(n)~\cite{Yin2025} & \SI{e-4}{}~\cite{Yin2025}\\ 
\midrule
\headcolor \multicolumn{4}{l}{\textbf{Layered Zipper Dielectrics}}
&\multicolumn{9}{c}{ } \\
\rowgreen Bi$_2$SeO$_5$ & 35~\cite{Khakbaz2025} & 22~\cite{Zhang2022} & GAA & yes & n & 2.4~\cite{Tang2025} & 0.45~\cite{Tang2025} & 300~\cite{Tang2025} & 0.6~\cite{Tang2025} & 69~\cite{Tang2025} & -0.1(n)~\cite{Tang2025} & 2000~\cite{Tang2025}\\ 
KAl$_3$Si$_3$O$_{10}$(OH)$_2$ & 8.1~\cite{Mohrmann2015} & -  & TG & no & n & 9~\cite{Zou2019} & 4.3~\cite{Zou2019} & 10~\cite{Zou2019}& 4~\cite{Zou2019} & 72~\cite{Zou2019}& +0.2(n)~\cite{Zou2019} & 0.1~\cite{Zou2019} \\ 
\midrule
\headcolor \multicolumn{4}{l}{\textbf{Native Oxides}}
&\multicolumn{9}{c}{ } \\
\roworange HfO$_2$ & 22~\cite{Kang2024, Fischetti2001} & 9~\cite{Li2019} & TG & yes & n & 10~\cite{Kang2024} & 1.7~\cite{Kang2024} & 10~\cite{Kang2024}& 1~\cite{Kang2024} & 62~\cite{Kang2024} & -0.6~\cite{Kang2024}  & 0.01~\cite{Kang2024} \\ 
ZrO$_2$ & 24~\cite{Fischetti2001} & 19~\cite{Jin2023} & TG & yes & n & 15~\cite{Jin2023} & 3~\cite{Jin2023} & 0.5~\cite{Jin2023} &3~\cite{Jin2023} & 90~\cite{Jin2023} & -1(n)~\cite{Jin2023} & 1~\cite{Jin2023} \\
Ta$_2$O$_5$ & 27.9~\cite{Martinez1974} & 15.5~\cite{Chamlagain2017a} & TG & yes & n & 13~\cite{Chamlagain2017a} & 3.3~\cite{Chamlagain2017a} & 0.5~\cite{Chamlagain2017a} & 1~\cite{Chamlagain2017a} & 70~\cite{Chamlagain2017a} & -0.4(n)~\cite{Chamlagain2017a} & 0.1~\cite{Chamlagain2017a}\\
\roworange c-Al$_2$O$_3$ & 12.5~\cite{Fischetti2001} & 5.5~\cite{Zeng2024} & TG & yes & n & 2~\cite{Zeng2024} &  1.4~\cite{Zeng2024} & 1.2~\cite{Zeng2024} & 0.5~\cite{Zeng2024} & 80~\cite{Zeng2024} & -0.2(n)~\cite{Zeng2024} & 0.04~\cite{Zeng2024} \\
\rowyellow Ga$_2$O$_3$ & 10.2~\cite{Passlack1994} & 22~\cite{Yi2024} & TG & no & n & 3.5~\cite{Yi2024} & 0.6~\cite{Yi2024} & 10~\cite{Yi2024} & 0.5~\cite{Yi2024} &  90~\cite{Yi2024} &-0.1(n)~\cite{Yi2024} & 0.4~\cite{Yi2024}\\ 
\midrule
\headcolor \multicolumn{4}{l}{\textbf{Fluorides}}
&\multicolumn{9}{c}{ } \\
\roworange CaF$_2$ & 6.8~\cite{Waldhoer2022}& 6.8~\cite{Waldhoer2022} & glob. BG & yes & n & 2.2~\cite{Illarionov2019} & 1.3~\cite{Illarionov2019} & 0.05~\cite{Illarionov2019} &1~\cite{Illarionov2019} & 110~\cite{Illarionov2019} & -0.5(n)~\cite{Illarionov2019} & 0.1~\cite{Illarionov2019}\\
\roworange LaF$_3$ & 14~\cite{Solomon1966} & 30~\cite{Meng2024} & loc. BG & yes & n &  100~\cite{Meng2024} & 3.5~\cite{Meng2024} & 5~\cite{Meng2024} & 3~\cite{Meng2024} & 70~\cite{Meng2024} &  -0.5(n)~\cite{Meng2024} & 0.01~\cite{Meng2024}\\
\midrule
\headcolor \multicolumn{4}{l}{\textbf{Transferred 3D Crystals}}
&\multicolumn{9}{c}{ } \\
\roworange SrTiO$_3$ & 330~\cite{Neville1972} & 30~\cite{Huang2022} & TG~\cite{Yang2022} & no & n & 15~\cite{Huang2022a} & 1.95~\cite{Huang2022a} &  0.1~\cite{Huang2022} & 0.9~\cite{Huang2022} & 72~\cite{Huang2022} & +0.3(n)~\cite{Huang2022} & 0.002~\cite{Huang2022}\\ 
\rowyellow Mn$_3$O$_4$ & 150~\cite{Yuan2025} & 75~\cite{Yuan2025} & TG & no & n & 15~\cite{Yuan2025} & 0.8~\cite{Yuan2025} & 0.2~\cite{Yuan2025} & 0.5~\cite{Yuan2025} & 95~\cite{Yuan2025} & -0.5(n)~\cite{Yuan2025} &0.003~\cite{Yuan2025}\\ 
\midrule
\headcolor \multicolumn{4}{l}{\textbf{Ferroelectric Insulators}}
&\multicolumn{9}{c}{ } \\
Hf$_{0.5}$Zr$_{0.5}$O$_2$ & - &- &  glob. BG & yes & n & 20~\cite{Mcguire2017} & - & 1~\cite{Mcguire2017} & 2~\cite{Mcguire2017}& 40~\cite{Mcguire2017}  & -1 (n)~\cite{Mcguire2017} & 0.1~\cite{Mcguire2017}\\ 
BaTiO$_3$ & - &- &  glob. BG & no & n & 48~\cite{Puebla2022} & - & 1.5~\cite{Puebla2022} & 10~\cite{Puebla2022} & -& -5(n)~\cite{Puebla2022}& 0.1~\cite{Puebla2022} \\ 
PbZr$_{0.5}$Ti$_{0.5}$O$_3$ & - &- &  glob BG. & yes & n & 500~\cite{Ko2016} & - & 0.2~\cite{Ko2016} & 3~\cite{Ko2016} &- & -1(n)~\cite{Ko2016} & 10~\cite{Ko2016} \\ 
Al$_{0.7}$Sc$_{0.3}$N & - &- &  glob. BG& yes& n& 100~\cite{Liu2021a}&- & 100~\cite{Liu2021a} & 30~\cite{Liu2021a} & - & -20 (n)~\cite{Liu2021a} &1000~\cite{Liu2021a} \\ 
\midrule
\headcolor \multicolumn{4}{l}{\textbf{Inorganic Molecular Crystals}}
&\multicolumn{9}{c}{ } \\
\roworange Sb$_2$O$_3$ & 11.5~\cite{Liu2021b} & - & TG & yes & n & 5~
\cite{Liu2021b} & 1.7~\cite{Liu2021b} & 0.002~\cite{Liu2021b} & 1~\cite{Liu2021b} & 75~\cite{Liu2021b} & -0.2(n)~\cite{Liu2021b} & 10~\cite{Liu2021b} \\
\bottomrule
\end{tabular}
\setcounter{table}{1}
\captionof{table}{Insulators that could serve as gate dielectrics for 2D FETs including their out-of-plane dielectric constants~($\varepsilon_\mathrm{r, \perp}$) and an overview over the performance of FETs that have been realized. Promising candidates with EOT$<\,$2nm are highlighted in orange, with EOT$<\,$1nm  in yellow, and with EOT$<\,$0.5nm in green. In the column geometry, the most advanced gate geometry that has been realized with this material is listed. The simplest geometry are FETs with a global back gate~(glob. BG), a bit more involved are those with a local back gate~(loc. BG), then top gated~(TG), double gated~(DG), and finally the most complex devices have a gate all around~(GAA). }
\label{tab:insulators}

\begin{tabular}{ccccccccccccc}
\toprule
\headcolor \textbf{Material} & $\varepsilon_\mathrm{r, \perp}$ bulk & $\varepsilon_\mathrm{r, \perp}$  at 0.5nm EOT & geom. & scal. & polar. & $t_\mathrm{ins}$ & EOT & $I_\mathrm{D,on}$ & $V_\mathrm{DD}$ & $SS$ & $V_\mathrm{th}$ & $I_\mathrm{G}$ at $V_\mathrm{DD}$\\
\headcolor - & $[1]$ & $[1]$ & - & - & - & $[\mathrm{nm}]$ & $[\mathrm{nm}]$ & [\textmu A\textmu$\mathrm{m}^{-1}$] & $[\mathrm{V}]$ & $[\mathrm{mV {dec.}^{-1}}]$ & $[\mathrm{V}]$ & $[\mathrm{mA{cm}^{-2}}]$\\
\midrule
\headcolor \multicolumn{4}{l}{\textbf{Amorphous Dielectrics}}
&\multicolumn{9}{c}{ } \\
TiOPc/Al$_2$O$_3$ & 12.5~\cite{Fischetti2001} & 5.5~\cite{Zeng2024} & TG & yes & n  & 5.3~\cite{Park2016f} & 2.9~\cite{Park2016f} & 80~\cite{Park2016f} &  1~\cite{Park2016f}&80~\cite{Park2016f}  & -0.5(n)~\cite{Park2016f} & 0.01~\cite{Park2016f} \\ 
 &  &  &  & &  p & 5.3~\cite{Park2016f} & 2.9~\cite{Park2016f} &  3~\cite{Park2016f} &  1~\cite{Park2016f} &  390~\cite{Park2016f} & +0.2(p)~\cite{Park2016f} & 0.01~\cite{Park2016f} \\ 
\roworange HfO$_2$ & 22~\cite{Kang2024, Fischetti2001} & 9~\cite{Li2019}& GAA & yes & n & 4~\cite{Mortelmans2024}& 1.7~\cite{Mortelmans2024} & 150~\cite{Mortelmans2024}& 0.8~\cite{Mortelmans2024} & 110~\cite{Mortelmans2024} & -0.1(n)~\cite{Mortelmans2024}& 6~\cite{Mortelmans2024}\\ 
\roworange  & & & & & p& 4~\cite{Mortelmans2024}& 1.7~\cite{Mortelmans2024} & 80~\cite{Mortelmans2024}& 2~\cite{Mortelmans2024} &160~\cite{Mortelmans2024} &-0.5(p)~\cite{Mortelmans2024}& 300~\cite{Mortelmans2024}\\
SiO$_2$ & 3.9 & 3.9 & glob. BG & yes & n & 30~\cite{Smithe2017a} & 30~\cite{Smithe2017a} & 60~\cite{Chou2020}& 10~\cite{Chou2020} & 80~\cite{Smithe2017a} & -3(n)~
\cite{Smithe2017a} & 0.2~\cite{Smithe2017a}\\ 
 & & &  TG & yes & p & 25~\cite{Ho2023} & 25~\cite{Ho2023} & 10~\cite{Ho2023} & 5~\cite{Ho2023} & 89~\cite{Ho2023} & -0.5(p)~\cite{Ho2023} & 0.01~\cite{Ho2023}\\ 
 ZrO$_2$ & 24~\cite{Fischetti2001} & 19~\cite{Zhao2022} & loc. BG & yes & n & 20~\cite{Zhao2022} & 4.1~\cite{Zhao2022} &  1~\cite{Zhao2022} &3~\cite{Zhao2022}& 70~\cite{Zhao2022} & +0.1(n)~\cite{Zhao2022} & 1~\cite{Zhao2022} \\
Y$_2$O$_3$ & 14~\cite{Onisawa1990} & 17.5~\cite{Wang2022} & TG & yes & n & 20~\cite{Wang2022} & 4.5~\cite{Wang2022} & 0.1~\cite{Wang2022} & 1.5~\cite{Wang2022} & 63~\cite{Wang2022} & -0.8(n)~\cite{Wang2022} & 0.4~\cite{Wang2022} \\ 
\roworange Er$_2$O$_3$ & 12~\cite{Losurdo2007} & 15~\cite{Uchiyama2023} & TG & yes & n  & 5~\cite{Uchiyama2023} & 1.3~\cite{Uchiyama2023}  & 0.01~\cite{Uchiyama2023}  & 2~\cite{Uchiyama2023} & 90~\cite{Uchiyama2023} & -0.8~\cite{Uchiyama2023} & 0.01~\cite{Uchiyama2023}\\ 
AlN & 8.5~\cite{Bhattacharjee2017} & - & glob. BG & yes & n & 150~\cite{Bhattacharjee2017} &  69~\cite{Bhattacharjee2017} & 60~\cite{Bhattacharjee2017}  &  3~\cite{Bhattacharjee2017} & 3000~\cite{Liu2021a} & -8~\cite{Bhattacharjee2017} & - \\ 
\headcolor \multicolumn{4}{l}{\textbf{Combined Gate Stacks}}
&\multicolumn{9}{c}{ } \\
\roworange PTCDA/HfO$_2$ & 22~\cite{Kang2024, Fischetti2001} & 9~\cite{Li2019}& TG & yes & n & 3~\cite{Li2019}& 1.3~\cite{Li2019} & 20~\cite{Li2019}& 0.5~\cite{Li2019} & 73~\cite{Li2019} & -1(n)~\cite{Li2019}& 0.05~\cite{Li2019}\\ 
\roworange  & & & & &  p& 3~\cite{Li2019}& 1.3~\cite{Li2019} & 0.01~\cite{Li2019}& 1~\cite{Li2019} &83~\cite{Li2019} &-0.6(p)~\cite{Li2019}& 0.003~\cite{Li2019}\\
\roworange AlO$_x$/HfO$_2$ & - & - & DG & yes & n & 5~\cite{Wu2021a} & 2~\cite{Wu2021a} & 50~\cite{Wu2021a} & 1~\cite{Wu2021a} & 100~\cite{Wu2021a} & +0.4(n)~\cite{Wu2021a} & 1000~\cite{Wu2021a} \\
\roworange hBN/Al$_2$O$_3$/HfO$_2$ & - & - & DG & no & n & 8~\cite{Lan2024} & 2~\cite{Lan2024} & 10~\cite{Lan2022} & 3~\cite{Lan2024} & 70~\cite{Lan2024} & -0.5 (n)~\cite{Lan2024}& 0.2~\cite{Lan2024} \\
\rowyellow hBN/HfO$_2$ & - & - & TG & yes & - & 1.7~\cite{Wang2024} & 0.55~\cite{Wang2024} & - & - & - & - & 0.001~\cite{Wang2024} \\
hBN/Al$_2$O$_3$ & - &- & loc. BG & no & n & 25.5~\cite{Piacentini2022} & 11~\cite{Piacentini2022} & 2~\cite{Piacentini2022} & 5~\cite{Piacentini2022} & 300~\cite{Piacentini2022} & -1.5 (n)~\cite{Piacentini2022} &  \SI{e-4}~\cite{Piacentini2022}\\
\rowyellow hAlN/HfO$_2$ &- &- & TG & yes & n & 3.7~\cite{Wang2023}& 0.7~\cite{Wang2023} & 1~\cite{Wang2023} & 3~\cite{Wang2023} & 93~\cite{Wang2023} & +1.2 (n)~\cite{Wang2023} &0.01~\cite{Wang2023}\\
CrOCl/hBN & - & - & loc. BG & no & p & 37~\cite{Guo2024}& 35~\cite{Guo2024}& 0.3~\cite{Guo2024}& 7~\cite{Guo2024}& - & -3.5(p)~\cite{Guo2024} & -\\ 
\rowyellow Al$_2$O$_3$/HfO$_2$ &  - &- & GAA & yes & n & 3~\cite{Chung2024} & 1~\cite{Chung2024} & 60~\cite{Chou2023} & 0.6~\cite{Chou2023} & 75~
\cite{Dorow2022}& -0.4(n)~\cite{Chou2023} & 10~\cite{Chou2023}\\
\rowyellow Al$_2$O$_3$/HfO$_2$ & - &- &DG& yes & n & 3~\cite{Jiang2023a} & 1~\cite{Jiang2023a} & 650~\cite{Jiang2023a} & 0.6~\cite{Jiang2023a} & 61~\cite{Jiang2023a} & +0.1(n)~\cite{Jiang2023a} & 1000~\cite{Jiang2023a}\\
\rowyellow Al$_2$O$_3$/HfO$_2$ & - &- &DG& yes & n & 2.9~\cite{Jiang2024} & 1~\cite{Jiang2024} & 400~\cite{Jiang2024} & 0.6~\cite{Jiang2024} & 63~\cite{Jiang2024} & +0.2(n)~\cite{Jiang2024} & 100~\cite{Jiang2024}\\ 
\roworange SiO$_2$/HfO$_2$ & - & - & TG & yes & n & 4.4~\cite{Ko2025a} & 2~\cite{Ko2025a} & 5~\cite{Ko2025a} & 1~\cite{Ko2025a} & 100~\cite{Ko2025a} & 0~\cite{Ko2025a} & 0.001~\cite{Ko2025a}\\
\roworange GdAlO$_x$/HfO$_2$ & - & - &DG & yes & n & 5.6~\cite{Wu2021a} & 2~\cite{Wu2021a} & 50~\cite{Wu2021a} & 1~\cite{Wu2021a} & 100~\cite{Wu2021a} & +0.5(n)~\cite{Wu2021a} &  100~\cite{Wu2021a} \\
\rowyellow Sb$_2$O$_3$/HfO$_2$ & -& - &TG & yes & n & 3~\cite{Xu2023} & 0.67~\cite{Xu2023} & 200~\cite{Xu2023} & 2~\cite{Xu2023} & 69~\cite{Xu2023} & -0.8(n)~\cite{Xu2023} & 0.125~\cite{Xu2023}\\
Ta$_2$O$_5$/HfO$_2$ & - & - &DG & yes & n & 8.5~\cite{Lan2023} & 2.6~\cite{Lan2023} & 150~\cite{Lan2023}  & 0.6~\cite{Lan2023} & 70~\cite{Lan2023} & -0.8(n)~\cite{Lan2023} & 10~\cite{Lan2023} \\
\bottomrule
\end{tabular}
\setcounter{table}{1}
\captionof{table}{[CONTINUED] Insulators that could serve as gate dielectrics for 2D FETs including their out-of-plane dielectric constants~($\varepsilon_\mathrm{r, \perp}$) and an overview over the performance of FETs that have been realized. Promising candidates with EOT$<\,$2nm are highlighted in orange, with EOT$<\,$1nm  in yellow, and with EOT$<\,$0.5nm in green. In the column geometry, the most advanced gate geometry that has been realized with this material is listed. The simplest geometry are FETs with a global back gate~(glob. BG), a bit more involved are those with a local back gate~(loc. BG), then top gated~(TG), double gated~(DG), and finally the most complex devices have a gate all around~(GAA). }

\end{landscape}
\clearpage
}
\newpage

\subsection{Methods for Transferring Gate Dielectrics}
\label{subsec:SI:transfer}
Most transfer methods are based on a polymer layer as a carrier scaffold, typically PMMA or PS in a wet transfer or PDMS in a dry transfer process~\cite{Frisenda2018}.
However, transferred dielectrics are usually contaminated by particles from the polymer scaffold~\cite{Tilmann2023} and the growth substrate.
In order to avoid contamination, a polymer-free transfer method based on silicon nitride cantilevers was developed~\cite{Wang2023a}.
An alternative approach is the vdW pick-up transfer method, where the entire stack of layered materials is formed by subsequently picking up one layer after the other with a PDMS stamp.
Typically, hBN is used as the top layer, which then either serves as encapsulation or top gate dielectric~\cite{Frisenda2018}.
However, this method is delicate and easily creates cracks in the 2D material, while shifting the contamination towards the top layered material. Hence, for the formation of top gate dielectrics this process can work, as the organic contamination will be at the interface between the insulator and the metal gate -- an interface where a certain degree of contamination appears acceptable.
For the bottom gate, the organic residues will be located at the critical insulator to channel interface, thereby severely degrading the performance. 
Usually, layered dielectrics like hBN~\cite{Roy2014, Lan2024}, manganese aluminum sulfide~(MnAl$_2$S$_4$, MAS)~\cite{Xu2022}, lanthanum oxybromide~(LaOBr)~\cite{Jiang2023, Soll2024} or gadolinium oxide~(Gd$_2$O$_5$)~\cite{Yin2025} are transferred on top of TMD channels.
In addition, also three dimensional (3D) crystals, for example perovskites like strontium titanate~(SrTiO$_3$) can be transferred on top of 2D semiconductors. 
In this case, the target perovskite films have to be epitaxially grown on a lattice-matched oxide sacrificial layer~\cite{Lu2016, Yang2022} using for example MBE, or pulsed laser deposition~(PLD). Next, a polymer coating is deposited on top of the perovskite and the sacrificial oxide is selectively etched in water to release the perovskite film into a thin freestanding membrane which can be transferred~\cite{Lu2016}. This has the disadvantage that the release process creates an interface full of dangling bonds which then forms the critical semiconductor to gate insulator interface. 
Another method to transfer 3D crystals that in principle offers better interface quality is remote epitaxy where a thin van der Waals crystal, e.g. graphene or hBN, serves as a growth template for an epitaxial layer and acts at the same time as a release layer, along which the thin film is peeled off and subsequently transferred~\cite{Kim2022, Park2024}. 

In recent years scalable transfer methods~\cite{Schram2022, Ghosh2023} and tools~\cite{Phommahaxay2019} for CVD grown TMDs have been developed. This scalable transfer process relies on the bonding of the transferrable layer on a glass carrier in an intermediate step, before the layer is transferred onto the target substrate and the glass carrier is laser de-bonded~\cite{Phommahaxay2019, Ghosh2024}.
Although this process was demonstrated for transferring TMDs (mostly monolayers) on the 300mm scale, it has never been used to transfer dielectrics.
Layered dielectrics are typically few-nanometers-thick and are therefore mechanically stronger than 2D channels, which may facilitate the transfer process and lead to fewer pinholes and cracks~\cite{Shen2021a}.
Recently, a wafer-scale transfer of ALD-grown amorphous Al$_2$O$_3$ and HfO$_2$ layers with thicknesses in the range from \SI{3}{nm} to \SI{20}{nm} has been demonstrated~\cite{Lu2023}.
Here a thin polyvinyl alcohol~(PVA) layer of \SI{9}{nm} is spin coated on silicon, depositing the ALD oxide on top and applying a thermal release tape. This stack is peeled off from the sacrificial silicon, the PVA is etched with an oxygen plasma and the oxide is transferred onto a 2D semiconductor~\cite{Lu2023}.
The 2D FET based on transferred ALD oxides showed good performance, even though the transfer introduced strain and organic contamination, causing a high device to device variability~\cite{Lu2023}. 

\subsection{Best Practices for Capacitance Measurements}
\label{subsec:SI:capmeas}
In the following, we provide ``best'' practices to reliably evaluate the CET of scaled gate stacks, see Figure~\ref{fig:perfeval}(c).
First, the "dual gate $V_\mathrm{th}$ leverage method", sometimes used to extract the top gate CET from the ratio of the top gate CET~(unknown) to bottom CET~(known)~\cite{Lee2022, Xu2023, Ko2025, Ko2025a}, can be prone to errors due to an arbitrary selection of the voltage and current ranges. Therefore, these measurements should be complemented by $C_\mathrm{G}(V_\mathrm{G})$ measurements on capacitors.
These capacitors for ultra-thin 2D channels should not be integrated like the vertical two-terminal MIS structures for bulk semiconductors, because a global bottom contact to the 2D channel would perturb the Fermi level modulation. Instead, a lateral design is needed, as shown in Figure~\ref{fig:perfeval}(d), which can be integrated in the same flow as 2D top gated FETs.
The multi-finger structure ensures that the channel length remains sufficiently short to enable admittance measurements at $\sim1\,$\textmu$\mathrm{m}$, see Subsection~\ref{subsec:interface}, while still allowing large gated areas~\cite{Gaur2019, Gaur2020}. 
Capacitors with varying gated areas ranging from e.g. $\SI{10}{}$\,\textmu $\mathrm{m^2}$ to $\SI{1000}{}$\,\textmu$\mathrm{m^2}$ should be considered.
$C_\mathrm{G}(V_\mathrm{G})$ measurements should be performed from the depletion into accumulation regime and the proper area normalization $C_\mathrm{G, meas}^{\prime}=C_\mathrm{G, meas}/A$  should be verified when determining the CET~\cite{Li2019, Wu2021a, Huang2022a, Zeng2024}.
In addition, it is essential to evaluate the capacitance over a wide frequency range, e.g., $f\in \left[\SI{1}{kHz}, \SI{1}{MHz}\right]$, as for insulator with sizable densities of mobile ions or charged point defects, the formation of an electric double layer can lead to an overestimation of the CET for measurements at low frequencies~\cite{Meng2024}.
For many novel gate dielectrics, fabricating capacitors with areas of at least $\SI{10}{}$\,\textmu $\mathrm{m^2}$ is challenging, due to frequently observed large densities of pinholes. 
However, these large areas are required to reach capacitances larger than \SI{0.1}{pF} that can be reliably characterized.  
When comparing the capacitance of dual gated 2D FETs to the CFET IRDS targets, the area normalization needs special attention.
As for silicon nanosheets the effective total width is calculated using the nanosheet perimeter, which corresponds to the double gate equivalent in 2D FETs~($A = 2\times W_\mathrm{sheet} \times L$)~\cite{Wu2021a, Lin2021}. In addition, CET can be verified using Hall-effect measurements, by evaluating the slope of the carrier density $n_\mathrm{S}$ as a function of $V_\mathrm{G}$. The carrier density $n_\mathrm{S}(V_\mathrm{G})$ extracted from Hall measurements can also provide insights into area normalization, and whether single or double gated charge centroid assumptions are justified.
Overall, reaching a CET of \SI{0.9}{nm} and an EOT of about \SI{0.5}{nm} are stringent targets, that, to the best of our knowledge, have only been met for the gate insulator Bi$_2$SeO$_5$~\cite{Tang2025} at an EOT of \SI{0.45}{nm}, see Table~\ref{tab:insulators}, although these are prototypes whose variability is far from IRDS standards. 
In particular, for gate stacks with a small $t_\mathrm{ins}$ of a few nm it is challenging to maintain sufficiently small leakage currents, as there is a fundamental trade-off between achieving good capacitive gate control and small gate leakage currents.

\subsection{Methodology for Leakage Current Calculation}
\label{subsec:SI:method_leak}
In the Tsu-Esaki model, the leakage current density $I_\mathrm{G}$ is calculated by integrating a Wentzel-Kramers-Brillouin~(WKB) factor over the entire conduction band.
The WKB factor exponentially depends on the insulator thickness $t_\mathrm{ins}$, the tunnel mass $m_\mathrm{tun}$, the applied gate bias $V_\mathrm{G}$ and the energy barrier charge carriers need to overcome $q \Phi$, with the elementary charge $q$,
\begin{equation*}
I_\mathrm{G} \propto \int_E \mathrm{exp}\left(-\frac{t_\mathrm{ins}}{V_\mathrm{G}} \sqrt{ m_\mathrm{tun} \left( q \Phi - E\right)^{3}} \right) \mathrm{d}E.
\end{equation*}
In consequence, the gate leakage $I_\mathrm{G}$ depends exponentially on $t_\mathrm{ins}$.
All material constants that serve as input parameters for the model are shown in the band diagram in Figure~\ref{fig:perfcalc}(b) and in Table~\ref{tab:materials}, including the energy barriers $q \Phi$ given by the band offsets for electrons and holes, along with the tunnel masses $m_\mathrm{tun}$ and the dielectric constants.
The band offsets can either be calculated based on density functional theory~(DFT)~\cite{Osanloo2021, Khakbaz2025} or experimentally determined using photoemission spectroscopy~\cite{Bersch2008, McDonnell2013}.
It should be noted that the electron affinity, $\chi$, is a surface property of the material, extracted from a heterostructure of the dielectric interfaced with a semiconductor, while the band gap is typically calculated for the bulk material~\cite{Khakbaz2025}.
The tunnel masses must be calculated based on the complex band structure of the insulator, as eigenstates of the insulator with a complex $k$ vector correspond to evanescent states in the gap.
The incoming electron wave function decays and its transmission probability and thus its tunneling mass is related to its decay, governed by its complex $k$ vector~\cite{Sacconi2007}. 
Since $I_\mathrm{G}$ of different insulators must be compared at the same CET,
$I_\mathrm{G}$ depends indirectly on the dielectric constant $\varepsilon_\mathrm{ins}$.
The dielectric constant can be calculated using DFT in combination with the Berry phase approach~\cite{Laturia2018, Khakbaz2025} or measured in a $C_\mathrm{G}\left(V_\mathrm{G}\right)$ analysis~\cite{Li2019, Wu2021a}, see Section~\ref{sec:performance}\ref{subsec:capacitance}.
In the calculations a vdW gap of \SI{0.3}{nm} with a dielectric constant of 2 was considered for all materials except for Bi$_2$O$_2$Se/Bi$_2$SeO$_5$ which does not have a vdW gap.
Also, it should be noted that the calculated leakage currents shown in Fig.~\ref{fig:perfcalc}~(a) are the sum of the electron tunneling current and the hole tunneling current. For example, for MoS$_2$/SiO$_2$ and MoS$_2$/hBN the hole tunneling current dominates also for the n-type for gate biases up to about \SI{0.6}{V} where the overall leakage shows a small dip, as hole and electron tunneling currents cancel each other out in a narrow gate bias region before the electron tunneling current dominates. 

\begin{table}[!ht]
\begin{tabular}{ccccccccc}
\toprule
Material & $\varepsilon_{r, \perp}$ bulk & $\varepsilon_{r, \perp}$ at 0.5nm EOT & $t_\mathrm{ins}$ at 0.5nm EOT & $t_\mathrm{DL}$ & $E_\mathrm{G}$ & $\chi$ & $m_\mathrm{tun,e}$ & $m_\mathrm{tun,h}$\\
 &  $[1]$ & $[1]$ &  $[\mathrm{nm}]$ & $[\mathrm{nm}]$ & $[\mathrm{eV}]$ & $[\mathrm{eV}]$ & $[1]$ & $[1]$ \\
 \midrule 
 hBN & 5.1~\cite{Geick1966} & 3.5~\cite{Laturia2018} &  0.6 (2 layers) ~\cite{Laturia2018} & - &  5.95~\cite{Cassabois2016} & 0.96~\cite{Haastrup2018} & 0.5~\cite{Britnell2012a} & 0.5~\cite{Britnell2012a} \\
 Bi$_2$SeO$_5$ & 35~\cite{Khakbaz2025} & 22~\cite{Zhang2022} & 2.4 (3 layers) ~\cite{Tang2025} & - & 3.16~\cite{Khakbaz2025} & 2.29~\cite{Khakbaz2025}  & - & - \\
 CaF$_2$ & 6.8~\cite{Waldhoer2022} & 6.8~\cite{Waldhoer2022} & 0.9 &  - &  11.8~\cite{Chen2022} & 0.22~\cite{Chen2022} & 1~\cite{Vexler2009} & - \\
 SrTiO$_3$ & 330~\cite{Neville1972} & 30~\cite{Huang2022} & 2.85 & 0.5~\cite{Hwang2002} &  3.3~\cite{Chambers2001} & 3.4~\cite{Chambers2019} & 0.1~\cite{Lee2009} & - \\
 Al$_2$O$_3$ & 12.5~\cite{Fischetti2001} & 5.5~\cite{Zeng2024} & 0.7 & - & 5.5~\cite{Dicks2019}& 1.95~\cite{Dicks2019}& 0.35~\cite{Yeo2002} &-  \\ 
 HfO$_2$ & 22~\cite{Fischetti2001, Kang2024} & 9~\cite{Li2019} & 1.2 & - &5.7~\cite{Strand2022} & 2.4~\cite{Strand2022} & 0.11~\cite{Monaghan2009, Zhu2002} & - \\ 
 SiO$_2$ & 3.9 & 3.9 & 0.5 & - & 9 & 0.9 & 0.42~\cite{Vexler2005}  & 0.33~\cite{Vexler2005} \\ 
 \midrule 
 MoS$_2$ & 7.6~\cite{Frindt1963} & 6.4~\cite{Laturia2018} & 0.65 (1 layer) & - &   2.18~\cite{Hill2016}  & 3.9~\cite{Haastrup2018} & - & -\\
 Bi$_2$O$_2$Se & 99~\cite{Khakbaz2025} &  - & - & -&1.04~\cite{Khakbaz2025} & 4.15~\cite{Khakbaz2025} &- & -\\
 WSe$_2$ & - & 7.5~\cite{Laturia2018} & 0.65 (1 layer) & - & 2.08~\cite{Chiu2015} & 3.25~\cite{Chiu2015}&- &-\\
 \midrule
 Au & - & - & - & - & - & 5.1~\cite{Riviere1966} &- &-\\
\bottomrule
\end{tabular}
\caption{Material properties of selected gate dielectrics, 2D semiconductors and metals.}
\label{tab:materials}
\end{table}

\subsection{Methodology for Mobility Calculation}
\label{subsec:SI:method_mobility}
In order to understand the effect of the surrounding insulators on 2D semiconductors, we performed transport  calculations for monolayer MoS$_2$ sandwiched between two gate insulators in a double-gated configuration.~\cite{Gopalan2022}. 
We calculated the mobility by numerically solving the Linearized Boltzmann Transport Equation~(LBTE), that fully takes into account the anisotropy of the electronic band structure as well as scattering mechanisms. 
In particular, three different scattering mechanisms have been included, intra- and inter-valley intrinsic phonons in MoS$_2$, SO phonons originating from the dielectrics and charged impurities. The SO phonon modes of various dielectrics were calculated using density functional perturbation theory~(DFPT).

\subsection{Best Practices for Interface Trap Density Measurements}
\label{subsec:SI:measinttraps}
Suitable measurement methods to evaluate the interface trap density include the analysis of admittance measurements~\cite{Gaur2019, Reato2025} and current transient spectroscopy~\cite{Taniguchi2018, Yang2024}. 
As a best practice example, for analyzing the interface quality of an insulator with the 2D channel $C_\mathrm{G}\left(V_\mathrm{G}\right)$ measurements need to be performed on multi-finger MIS capacitor structures with gated areas on the order of 100-1000\textmu $\mathrm{m^{2}}$, see Figure~\ref{fig:perfeval}~(d).
As described in Subsection~\ref{sec:performance}\ref{subsec:capacitance} edge-MIS capacitors are required for $C_\mathrm{G}\left(V_\mathrm{G}\right)$ measurements~\cite{Gaur2020}. 
Since the charge is injected laterally from the S/D fingers the lateral flow of carriers creates an RC effect and the frequency dispersion in the $C_\mathrm{G}\left(V_\mathrm{G}\right)$ response could be confused with $D_\mathrm{it}$.
This resistance contribution can be minimized with the multi-finger capacitor design, where $L_\mathrm{TG}$ should be kept $\leq1$\textmu$\mathrm{m}$ for channel mobilities in the range $10-\SI{40}{cm^2V^{-1}s^{-1}}$. 
Measurements at different frequencies $\SI{1}{kHz}-\SI{1}{MHz}$ and at different temperatures $\SI{4}{K}-\SI{300}{K}$ enable the extraction of the $D_\mathrm{it}\left(E\right)$ profile by the conductance method~\cite{Gaur2019, Gaur2020, Gaur2020a}.
An example is shown in Figure~\ref{fig:perfeval}(e) for different stacks on MoS$_2$ and WS$_2$ channels with a TMA soak interlayer~\cite{Mootheri2021}. 
As the MoS$_2$ channels are n-type only, the $D_\mathrm{it}\left(E\right)$ profile can only be accessed close to the conduction band edge.
The WS$_2$ channels are ambipolar, enabling profiling along the full band gap. 
For both channels, the $D_\mathrm{it}\left(E\right)$ near mid-gap are close to the targeted $\SI{e12}{cm^{-2}eV^{-1}}$, but near the band edges, the $D_\mathrm{it}$ increases exponentially to $\SI{e14}{cm^{-2}eV^{-1}}$.
This causes a stretch-out of the transfer characteristics, preventing low-power operation.

\subsection{Hysteresis and BTI Measurements}
\label{subsec:SI:meashystBTI}
Most frequently, a positive hysteresis is measured~($\Delta V_\mathrm{H}>0$), caused by charge trapping in the insulator from the 2D channel. 
At the same time, insulator traps close to the gate can also change their occupancy during a sweep. As they become discharged as $V_\mathrm{G}$ is swept up, the resulting $\Delta V_\mathrm{H}$ will be negative and result in a negative hysteresis. In addition to charge trapping, also the drift and diffusion of mobile ions can contribute to a negative hysteresis~($\Delta V_\mathrm{H}<0$)~\cite{Knobloch2023a, Karl2025}. For example, sodium and potassium ions can be introduced from seeding promoters for CVD of 2D materials~\cite{Lee2013d, Zhu2023a}. Another possible cause for a negative hysteresis would be the switching of the polarization of a ferroelectric layer in the gate stack~\cite{Mcguire2017, Cao2020}. For p-type FETs the sign convention of $\Delta V_\mathrm{H}$ changes. As all of these mechanisms have different temperature and sweep time dependencies, measurements of the hysteresis over many orders of magnitude in sweep times and a wide range of temperatures can be used to determine the root cause of the observed hysteresis~\cite{Karl2025}, see Figure~\ref{fig:perfeval}(c).

Usually, the measured BTI response is strongly asymmetric, meaning that it takes much longer for $\Delta V_\mathrm{th}$ to recover than to build up. Again, the most important contribution to observed BTI drifts comes from charge trapping at border traps at the channel side of the gate stack~\cite{Huard2006, Grasser2012}. 
In n-type FETs, usually a $V_\mathrm{H}$ at $V_\mathrm{DD}$ or above is chosen and a $V_\mathrm{L}$ close to $V_\mathrm{th}$, thus $V_\mathrm{H}>V_\mathrm{L}$, resulting in positive BTI~(PBTI).
Typically, the observed threshold voltage shifts $\Delta V_\mathrm{th}$ during PBTI are positive, which is referred to as normal BTI, whereas negative $\Delta V_\mathrm{th}$ shifts are termed anomalous.
In an analogous way, in p-type FETs, $-V_\mathrm{H}$ is larger than $-V_\mathrm{L}$, resulting in negative BTI~(NBTI) with negative $\Delta V_\mathrm{th}$ drifts.
In a best case scenario, the  $\Delta V_\mathrm{th}$ drifts should be monitored at geometrically increasing intervals during both the stress and recovery phases, for example using very fast $V_\mathrm{G}$ sweeps~\cite{Dorow2022, Provias2023}, see Figure~\ref{fig:perfeval}(c). However, especially during recovery some signal is lost during the fast $V_\mathrm{G}$ sweep read-outs, thus for capturing also the response of fast traps, the drain current at $V_\mathrm{th}$ is recorded and later on mapped to $\Delta V_\mathrm{th}$ shifts~\cite{Kaczer2008}. 

\subsection{Dielectric Breakdown Measurements}
\label{subsec:SI:breakdownmeas}
In the best case, TDDB should be analyzed based on constant voltage stress~(CVS) measurements, where a constant voltage stress is applied for a long time span until the insulator breaks.
As BD is a statistical phenomenon, CVS measurements need to be performed on large number of MIS capacitors and the onset of breakdown is then typically evaluated statistically in a Weibull distribution~\cite{Ranjan2019, Strand2022}.
The statistical nature of BD makes the evaluation extremely time consuming, especially for novel dielectrics, as many CVS measurements are required on many samples that by definition will all be destroyed after the measurements. 
A more practical approach to characterize BD are ramped voltage stress~(RVS) experiments, where the gate voltage is gradually increased until the insulator breaks~\cite{Hattori2015, Ranjan2019, Yu2020}. For lifetime extrapolations, RVS TDDB data can subsequently be converted to CVS TDDB statistics~\cite{Kerber2007}. Ideally, TDDB should be analyzed for multiple MIS capacitors of varying areas, and insulator thicknesses evaluating both the ramp-rate dependent breakdown voltage~($V_\mathrm{BD}$) as well as its Weibull slope~($\beta_\mathrm{V_{BD}}$), see Figure~\ref{fig:perfeval}(c).

\subsection{Experimental Leakage Currents through hBN}
\label{subsec:SI:hBNleakage}
Reference~\cite{Britnell2012a} reported that hBN with thicknesses of 1L, 2L 3L and 4L shows leakage currents of of $\SI{21.5}{kAcm^{-2}}, \SI{9.5}{kAcm^{-2}}, \SI{210}{Acm^{-2}}$ and $\SI{7.9}{Acm^{-2}}$ (respectively), at a bias of \SI{0.6}{V}. However, these values have been measured in graphite/hBN/graphite capacitors with areas of 2-10\textmu$\mathrm{m}^2$  by showing one forward RVS measured in one device per thickness. This is particularly important because many other studies have reported leakage currents down to \SI{1}{mAcm^{-2}}-\SI{0.01}{mAcm^{-2}} at a bias of \SI{0.6}{V}~\cite{Wang2018, Wu2019a}, although we consider such values to be impossible and those studies affected by the presence of moisture and/or polymer residues. Therefore, these values need to be confirmed statistically and in devices with identical sizes. Furthermore, it is critical to analyse hBN thicknesses between 4 and 15 layers, not only to understand the quantum tunnelling current across them, but also to investigate what hBN thickness leads to acceptable gate leakage currents below \SI{0.8}{Acm^{-2}} at a bias of \SI{0.6}{V}.





\end{document}